\documentclass[aps,prd,twocolumn,amssymb,eqsecnum,showpacs,nofootinbib]{revtex4}
\usepackage{graphicx}
\usepackage{amsmath}
\usepackage{amssymb}
\usepackage{bm}
 \usepackage{color}

\def\nn{\nonumber}
\def\be{\begin{equation}}
\def\ee{\end{equation}}
\def\ba{\begin{eqnarray}}
\def\ea{\end{eqnarray}}

\def\mpl{m_{\rm p}}
  \def\be{\begin{equation}}
\def\ee{\end{equation}}
 \def\bi{\begin{itemize}}
 \def\ei{\end{itemize}}
  \def\ben{\begin{enumerate}}
\def\een{\end{enumerate}}
  \def\bt{\begin{tabular}}
\def\et{\end{tabular}}
\def\bc{\begin{center}}
\def\ec{\end{center}}

\def\bea{\begin{eqnarray}}
\def\eea{\end{eqnarray}}

\newcommand{\bes}{\begin{subequations}}
\newcommand{\ees}{\end{subequations}}

\begin{document}

\title{Adiabatic instability in coupled dark energy-dark matter models}
\author{Rachel Bean$^{1}$}
\author{\'Eanna \'E. Flanagan$^{1,2}$}
\author{Mark Trodden$^{3}$}

\affiliation{$^{1}$ Department of Astronomy, Cornell University, Ithaca, NY 14853, USA}
\affiliation{$^{2}$Laboratory for Elementary Particle Physics, Cornell University, Ithaca, NY 14853, USA.}
\affiliation{$^{3}$Department of Physics, Syracuse University, Syracuse, NY 13244, USA}

%
%
\newcount\hh
\newcount\mm
\mm=\time
\hh=\time
\divide\hh by 60
\divide\mm by 60
\multiply\mm by 60
\mm=-\mm
\advance\mm by \time
\def\hhmm{\number\hh:\ifnum\mm<10{}0\fi\number\mm}


\date{\today}

\begin{abstract}

We consider theories in which there exists a nontrivial coupling between the
dark matter sector and the sector responsible for the acceleration of the universe.
Such theories can possess an adiabatic regime in which the
quintessence field always sits at the minimum of its effective
potential, which is set by the local dark matter density.
We show that if the coupling strength is much larger than
gravitational, then the adiabatic regime is always
subject to an instability.  The instability, which can also be thought of as a type
of Jeans instability, is characterized by a negative sound speed squared of
an effective coupled dark matter/dark energy fluid, and results in the
exponential growth of small scale modes.
We discuss the role of the instability in specific coupled CDM and Mass Varying Neutrino (MaVaN) models
of dark energy, and clarify for these theories the regimes in which the instability
can be evaded due to non-adiabaticity or weak coupling.

\end{abstract}
\maketitle

\section{Introduction}
\label{intro}

In order for our cosmological models to provide an accurate fit to all
current observational data, it is necessary to postulate two dramatic
augmentations beyond the minimalist assumption of baryonic matter
interacting gravitationally through Einstein's equations. The first
assumption is that there must exist either new gravitational dynamics
or a new component of the cosmic energy budget -- {\it dark matter} --
that allows structure to form and accounts for weak lensing and
galactic rotation curves. The second assumption is that a further
dynamical modification or energy component -- {\it dark energy} --
exists, driving late-time cosmic acceleration.

In the first case, recent results~\cite{Clowe:2006eq} have added
significant support to an explanation in terms of particulate
cold dark matter (CDM), rather than a modification of gravity. In the case of
cosmic acceleration, however, the data remains consistent with a
simple cosmological constant, with modifications to gravity, or with
dark energy as the correct explanation.

A logical possibility is that the two dark sectors -- dark matter and
dark energy -- interact with each other or with the visible sector of
the theory
\cite{Damour:1990,Carroll:1998zi,Amendola:1999er,Bean:2000zm,Bean:2001ys,
 Chiba:2003ir,Majerotto:2004ji,Das:2005yj,Lee:2006za,Kesden:2006zb}.
In fact, a number of models have been proposed that exploit this idea
to address, among other things, the coincidence problem
\cite{Bean:2000zm,Bean:2001ys}. Further,
there exist classes of modified gravity models which may either be
mapped to interacting dark energy models, or closely approximated by
them over a broad range of dynamical interest \cite{Chiba:2003ir,Tsujikawa:2007gd}.

There are concerns, however, about coupling these two rather
differently behaving sectors. One concern is the presence of dynamical
attractors that produce a cosmic expansion history significantly
different from $\Lambda CDM$; see, for example, Refs.\
\cite{Amendola:1999er,Amendola:2006mr,Agarwal:2007}.
Another specific example is
the possibility of
instabilities that are not present for the uncoupled system
\cite{Afshordi:2005ym,Kaplinghat:2006jk,Bjaelde:2007ki}.

In this paper we perform a careful analysis of the viability of
coupled dark energy-dark matter models.
We consider implications of
such coupled theories on cosmological scales, both in terms of
homogeneous background expansion and the growth of linear
perturbations. We focus, in particular, on a
class of models in which there exists an {\it adiabatic regime} in
which the dark energy field instantaneously tracks the minimum of its
effective potential, as explored by Das, Corasaniti and
Khoury\cite{Das:2005yj}.
We show that if the coupling strength
is much larger than gravitational, then the adiabatic regime is always
subject to an instability.  The instability is characterized by a negative sound speed squared of
an effective coupled dark matter/dark energy fluid, and results in the
exponential growth of small scale modes.
We analyze a number of different models, and show that for these
models the instability
strongly constrains the region in parameter space that is compatible with observations.
A short version of our results was given in the recent paper \cite{letter}.

We improve on previous investigations of this instability
\cite{Afshordi:2005ym,Kaplinghat:2006jk,Bjaelde:2007ki} in a number of
ways.  First, we show that the instability occurs only for coupling strengths that are strong
compared to gravitational coupling, and can be evaded at weaker
couplings; this point was missed all in previous work.
Second, we give a simple intuitive explanation of the instability
as a type of Jeans instability.  Normally, for cosmological perturbations, Hubble damping
converts the exponential growth of gravitationally unstable modes into power law growth.
However, here the Hubble damping is ineffective, due to the fact that
the effective Newton's constant for the interaction of dark matter
with itself is much larger than the Newton's constant governing the
background cosmology.  The result is the exponential growth of perturbations.
Finally, we generalize previous treatments to allow an arbitrary coupling between
the dark energy sector and the visible sector, in addition to the
coupling between dark energy and dark matter.

In more detail, we consider models which are characterized by a
function $\alpha_c(\phi)$ governing the interaction between CDM and a
quintessence type scalar field $\phi$, and a function $\alpha_b(\phi)$
governing the interaction between visible matter (baryons) and $\phi$.
The effective Newton's constants
for the interaction of CDM with itself (cc), the interaction of CDM
with baryons (cb), and the interaction of baryons with themselves (bb)
are (see Appendix \ref{sec:Geff})
\bes
\label{Newtons}
\bea
G_{cc} &=& G \left [1 + 2 \mpl^2 \alpha_c'(\phi)^2 \right], \\
G_{cb} &=& G \left [1 + 2 \mpl^2 \alpha_c'(\phi) \alpha_b'(\phi)\right], \\
G_{bb} &=& G \left [1 + 2 \mpl^2 \alpha_b'(\phi)^2 \right].
\eea
\ees
These Newton constants are those of the Einstein frame in the short
wavelength limit, and include the effect of the scalar interaction
mediated by $\phi$.  Here $\mpl$ is the Planck mass.

The adiabatic instability occurs when $G_{cc} \gg G$, or equivalently
$\mpl \alpha_c' \gg 1$, when the scalar coupling between dark matter
particles is strong compared to the tensor coupling\footnote{This
requirement is admittedly a fine tuning and unnatural from the
viewpoint of effective field theory.}.
In the present day Universe, this regime is excluded by observations
if we assume that perturbations to the cosmological background value of
$\phi$ are in the linear regime, so that the parameters
(\ref{Newtons}) are constants (see below).
First, tests of general relativity in the Solar System constrain $\mpl
\alpha_b'$ to be small compared to unity.  Second, observations of
tidal disruption of satellite galaxies of the Milky Way provide the
constraint\footnote{Only the combination (\ref{Mark}) of the Newton's constants
  can be constrained by observations of the gravitational interactions
  of baryons and dark matter, since once cannot separately measure the
  densities and the Newton's constants.} \cite{Kesden:2006zb}
\be
\left| \frac{G_{bc}}{\sqrt{G_{cc} G_{bb}}} -1 \right| \alt 0.02.
\label{Mark}
\ee
Combining these constraints excludes the regime $G_{cc} \gg G$.

Nevertheless, it is still of interest to explore the adiabatic
instability, for a number of reasons.
First, the argument given above assumes a single type of CDM particle.
In more complicated models with two different CDM species, one coupled
and one uncoupled, the constraint can be evaded (see Secs.\
\ref{sec:twocomponent} and \ref{sec:MaVaN} below).  Second, the Newton's
constants can evolve as a function of
redshift, via the dependence of $\phi$ on redshift,
so present day observations $(z \approx 0)$ do not exclude the
occurrence of the instability at high redshifts.  Third, it is useful
to have multiple independent constraints on coupled models.
Finally, it is not necessarily true that perturbations to the
cosmological background value of $\phi$ are in the linear regime.  In particular, it
has been claimed that observational constraints on $\mpl
\alpha_b'$ can be evaded in chameleon models\footnote{In particular,
chameleon models in the regime $\mpl \alpha_b' \gg 1$
have been extensively explored \cite{Mota:2006ed}.}, due the
dependence of the local effective Newton's constants (\ref{Newtons})
on $\phi$ and hence on the local matter density \cite{Khoury:2003aq,Khoury:2003rn}.

The structure of the paper is as follows. In the next section, we
introduce the general class of models and the framework in which we
work.
In section~\ref{adiabatic} we discuss the
adiabatic regime, giving both local and nonlocal conditions for its
applicability.  We also derive the effective equation of state for
these models in the adiabatic regime; as previously noted in Ref.\
\cite{Das:2005yj}, superacceleration is a generic feature of these models.
Section \ref{instability} discusses the adiabatic instability that can
arise in the adiabatic regime.  We discuss two complementary ways of
understanding the instability, a hydrodynamic viewpoint and a Jeans
instability viewpoint.  We also derive the range of lengthscales over
which the instability operates, which was incomplete in earlier
analyses \cite{Afshordi:2005ym,Kaplinghat:2006jk}.
In
section~\ref{examples} we apply our general analysis to some
well-known coupled models. We perform analytic and numerical analyses of the evolution of
the background cosmology and of perturbations in both coupled CDM,
chameleon \cite{Khoury:2003aq,Khoury:2003rn}
and mass varying neutrino (MaVaN) \cite{Fardon:2003eh,Kaplan:2004dq,Fardon:2005wc} models, and identify
the regimes in which they are subject to the instability. We conclude in~\ref{conclusions} by summarizing
our findings and discussing their implications.

In appendix~\ref{visible} we generalize the derivation of the
instability to include baryonic matter, and in appendix~\ref{kinetic}
we generalize the derivation of the instability to regimes where
CDM cannot be treated as a pressureless fluid but instead must be
treated in terms of kinetic theory.
Finally, in appendix~\ref{sec:Geff} we derive and describe the Jeans
instability viewpoint on the instability.

A note on conventions: throughout this paper we use a metric signature
(-,+,+,+), and we define the reduced Planck mass by $\mpl \equiv
(8\pi G)^{-1/2} $.

\section{Models with a coupling between dark energy and dark matter}
\label{sec:web}

\subsection{General class of models}

We begin from the following action
\bea
S &=& S[g_{ab},\phi,\Psi_{\rm j}] \nonumber \\
&=& \int d^4x\sqrt{-g}
\left[ \frac{1}{2} \mpl^2 R
-\frac{1}{2} (\nabla \phi)^2
 - V(\phi)
\right] \nonumber \\
&&+ \sum_{\rm j} S_{\rm j}[e^{2 \alpha_{\rm j}(\phi)} g_{ab}, \Psi_{\rm j}],
\label{action0}
\eea
where $g_{ab}$ is the Einstein frame metric, $\phi$ is a scalar field
which acts as dark energy, and $\Psi_{\rm j}$ are the matter fields.
The functions $\alpha_{\rm j}(\phi)$ are coupling functions that determine the strength
of the coupling of the jth matter sector to the scalar field.
The matter sectors include cold dark matter with coupling function
$\alpha_c(\phi)$, and baryons with coupling function $\alpha_b(\phi)$.

This general action encapsulates many models studied in the
literature.
The case
of equal coupling to dark and visible matter,
$\alpha_{\rm j}(\phi) = \alpha(\phi)$ for all j\footnote{A more
precise characterization of equal coupling is $\alpha_{\rm j}'(\phi)
= \alpha'(\phi)$ for all j, since any constant term in $\alpha_{\rm
  j}(\phi)$ can be absorbed into a rescaling of all dimensionful
parameters in the action $S_{\rm j}$.},
corresponds to
scalar-tensor theories of gravity which have been extensively studied,
recently under the name of ``coupled quintessence''
\cite{Amendola:1999er}.
This class of theories includes the $f(R)$ modified gravity theories
\cite{Carroll:2003wy}.
Several authors have considered the case $\alpha_b(\phi)=0$, in which
the dark energy couples only to dark matter.
We note that this choice yields a
microphysical model
for Modified-Source Gravity~\cite{Carroll:2006jn}
in the adiabatic regime discussed in Sec.\ \ref{adiabatic} below
(in the approximation where one considers only a dark matter source
and neglects baryons).

There is good theoretical motivation for considering nontrivial and different
coupling functions $\alpha_{\rm j}(\phi)$, since
this is a generic prediction
of string theory and of higher dimensional models.
In fact, typically the moduli and dilation fields of string theory must be
massive today, because for massless fields it is difficult to satisfy the observational
constraints [Solar System tests and fifth force experiments that require
 $d\alpha_b/d\phi \alt 10^{-2} \mpl^{-1}$, equivalence principle
tests that require matter coupling functions for
different matter sectors aside from dark matter to be
very nearly the same] in a natural way because
of loop corrections.
However dynamical dark energy models require massless or nearly
massless fields so one is forced to confront the naturalness issue.
One possible solution, suggested by Damour and Polyakov
\cite{Damour:1994ya}, is that there is an attractor mechanism under
cosmological evolution that drives the theory to be very close to
general relativity, $\alpha_{\rm j}^\prime \to 0$ for all $j$.
It is possible that this mechanism does not work
perfectly, and that there are residual deviations in the form of
nontrivial matter couplings.

In addition, while equivalence principle tests strongly constrain
differences between the coupling functions for different types
of visible matter, the corresponding constraints on dark matter are
much weaker, as first pointed out by Damour, Gibbons and Gundlach \cite{Damour:1990}.
So there is good motivation for exploring models with various couplings
to dark matter.

If the jth sector consists of a fermion of mass $m$, the
corresponding action in Eq.\ (\ref{action0}) can be written as
\begin{eqnarray}
&& S_{\rm j} =
\int d^4 x
\sqrt{- g} \bigg[ e^{p \alpha(\phi)} {\bar \Psi}_{\rm j} i \gamma^\mu \nabla_\mu
 \Psi_{\rm j} - e^{q \alpha(\phi) } m {\bar \Psi}_{\rm j} \Psi_{\rm
   j} \bigg], \nonumber\\
\label{eq:action10}
\end{eqnarray}
with $p=2$ and $q=3$.  Other possibilities for
$p$ and $q$ have been explored in the
literature.  For example, Farrar and Peebles \cite{Farrar:2003uw}
consider the case $p=0$, and Bean \cite{Bean:2001ys} considers the
case $p=q$.  All of these choices are equivalent\footnote{Up to
  redefining $\alpha$ by multiplying by a constant.} in the
nonrelativistic limit, when the fermion's rest masses dominate their
gravitational interactions.
The choice we make here is motivated by the fact that it satisfies the
equivalence principle within the jth sector, and also the fact that it
arises naturally from higher dimensional models.

The field equations resulting from the action (\ref{action0}) are
\bea
\mpl^2 G_{ab} &=& \nabla_a \phi \nabla_b \phi - \frac{1}{2} g_{ab}
(\nabla \phi)^2 - V(\phi) g_{ab} \nonumber \\
&&
+ \sum_{\rm j}e^{4 \alpha_{\rm j}(\phi)} \left[ ({\bar \rho}_{\rm j} + {\bar
    p}_{\rm j}) u_{{\rm j}\,a} u_{{\rm j}\,b} + {\bar
    p}_{\rm j} g_{ab} \right],\ \ \
\label{ee0}
\eea
and
\be
\nabla_a \nabla^a \phi - V'(\phi) = \sum_{\rm j} \alpha_{\rm j}'(\phi)
e^{4 \alpha_{\rm j}(\phi)}
({\bar \rho}_{\rm j} - 3 {\bar p}_{\rm j} ).
\label{eq:scalar10}
\ee
where the prime represents a derivative with respect to $\phi$. Here we treat the matter $\Psi_{\rm j}$ as a fluid with Jordan-frame density
${\bar \rho}_{\rm j}$ and pressure ${\bar p}_{\rm j}$ and with a
4-velocity $u_{{\rm j}\,a}$ normalized according to $g^{ab} u_{{\rm
    j}\,a} u_{{\rm j}\,b}=-1$ [see Appendix \ref{visible}].

For much of this paper we shall work in the approximation where we consider only the
gravitational effects of the dark matter and neglect the gravitational
effects of the baryons.  In this approximation, the sums over j in
Eqs.\ (\ref{ee0}) and (\ref{eq:scalar10}) reduce to a single term involving the dark
matter.  Dropping the index j for simplicity, and denoting the
CDM coupling function $\alpha_c(\phi)$ simply
by $\alpha(\phi)$, the field equations become
\bea
\mpl^2 G_{ab} &=& \nabla_a \phi \nabla_b \phi - \frac{1}{2} g_{ab}
(\nabla \phi)^2 - V(\phi) g_{ab} \nonumber \\
&& + e^{4 \alpha(\phi)}
\left[ ({\bar \rho} + {\bar p}) u_a u_b + {\bar p} g_{ab} \right],
\label{ee0d}
\eea
and
\be
\nabla_a \nabla^a \phi - V'(\phi) = \alpha'(\phi) e^{4 \alpha(\phi)}
({\bar \rho} - 3 {\bar p} ),
\label{eq:scalar10a}
\ee
These are are the standard equations for a
scalar-tensor cosmology.

For dark matter we have ${\bar
  p}=0$, and we define a rescaled density variable
\be
\rho \equiv e^{3 \alpha(\phi)} {\bar \rho}.
\label{newdensity}
\ee
With this new variable, we obtain
\be
\mpl^2 G_{ab} = \nabla_a \phi \nabla_b \phi - \frac{1}{2} g_{ab}
(\nabla \phi)^2 - V(\phi) g_{ab} + e^{\alpha(\phi)}
\rho u_a u_b,
\label{ee}
\ee
and
\be
\nabla_a \nabla^a \phi - V'(\phi) = \alpha'(\phi) e^{\alpha(\phi)}
\rho \ .
\label{eq:scalar1}
\ee
The scalar field equation (\ref{eq:scalar1}) can also be written as
\be
\nabla_a \nabla^a \phi - V_{\rm eff}'(\phi) = 0,
\label{eq:scalar2}
\ee
where the effective potential that includes the matter coupling is
\be
V_{\rm eff}(\phi) = V(\phi) + e^{\alpha(\phi)}
\rho.
\label{phieqn}
\ee
and the prime means derivative with respect to $\phi$ at fixed $\rho$.
The equations of motion for the fluid are [see Appendix \ref{visible}]
\be
\nabla_a ( \rho u^a) =0,
\label{eq:fluid1}
\ee
and
\be
u^b \nabla_b u^a  = - (g^{ab} + u^a u^b) \nabla_b \alpha \ .
\label{eq:fluid2}
\ee

Note that the equations of motion (\ref{ee}) -- (\ref{eq:fluid2}) of the theory
do not depend on the coupling function $\alpha_b(\phi)$ of the
scalar field to visible matter, in the approximation where we neglect
the baryons.
Our discussion of the instability in the following sections
will be valid for arbitrary $\alpha_b(\phi)$.
However we note that $\alpha_b(\phi)$ will
enter when we want to compare with observations, since the metric that
is measured is $e^{2 \alpha_b} g_{ab}$.  When discussing observations below we will
focus on the case $\alpha_b=0$, where the dark energy is coupled only to dark matter.


\section{The Adiabatic Regime}
\label{adiabatic}

The effective potential (\ref{phieqn}) governing the evolution of the
scalar field
is the sum of two terms, one arising from the original
potential $V(\phi)$, and the other arising from the coupling to the
energy density of the dark matter fluid. It is possible, for
appropriate choices of the potential and the coupling function, that
competition between these terms leads to a minimum of the effective
potential. Further, in some regimes, it may be possible for the
solution of the equation of motion for $\phi$ to
adiabatically\footnote{Note that throughout this paper, we use
``adiabatic'' in the sense of ``gradual change'', not in the thermodynamic
sense of ``at constant entropy''.} in the  track
the position of this minimum.
That is to say, the timescale or lengthscale for $\phi$ to adjust
itself to the changing position of the minimum of the effective potential may be short
compared to the timescale or lengthscale over which the background
density is changing.

This {\it adiabatic regime} has been previously discussed for spatial
variations of $\phi$ in the interior of massive bodies in the so-called Chameleon field models
\cite{Khoury:2003aq,Khoury:2003rn}, and for the time variation of
$\phi$ in a cosmological context by Refs.\ \cite{Kaplinghat:2006jk,Das:2005yj}.
It also has been studied for the specific case of $f(R)$ modified gravity models
\cite{Navarro:2006mw,Faulkner:2006ub,Hu:2007nk}, for which the action
in the Einstein frame is of the general form (\ref{action0}).
In this section we review the adiabatic regime in a general context and give a careful
derivation of its domain of validity.

The adiabatic approximation consists of (i) omitting the d'Alembertian
term in Eq.\ (\ref{eq:scalar2}), which gives an algebraic equation for $\phi$
that one can solve to obtain $\phi$ as a function of the density $\rho$;
(ii) omitting the terms involving the gradient of $\phi$ from the field
equation (\ref{ee}).  The resulting equations are the same as those of
Modified-Source gravity~ \cite{Carroll:2006jn}, in the approximation
where one considers only a dark matter source and neglects baryons.
The equations can be written as general relativity
coupled to a fluid with an effective energy density $\rho_{\rm eff}$
and effective pressure $p_{\rm eff}$:
\be
\mpl^2 G_{ab} =
\left[ (\rho_{\rm eff} + p_{\rm eff}) u_a u_b + p_{\rm eff} g_{ab} \right] \ ,
\label{ee1}
\ee
where
\be
\rho_{\rm eff}(\rho) = e^{\alpha[\phi_{\rm m}(\rho)]} \rho
+ V[\phi_{\rm m}(\rho)] \ ,
\label{rhoeff}
\ee
\be
p_{\rm eff}(\rho) = - V[\phi_{\rm m}(\rho)] \ .
\label{peff}
\ee
Here $\phi_{\rm m}(\rho)$ is the solution of the algebraic
equation
\be
V_{\rm eff}'(\phi)= V'(\phi) + \alpha'(\phi) e^{\alpha(\phi)} \rho =0
\label{eq:alg}
\ee
for $\phi$.
Eliminating $\rho$ between Eqs.\ (\ref{rhoeff}) and (\ref{peff}) gives
the equation of state $p_{\rm eff} = p_{\rm eff}(\rho_{\rm eff})$.

In the adiabatic regime, the matter and scalar field are tightly
coupled together and evolve as one effective fluid.  By taking the
divergence of the field equation (\ref{ee1}), we see that this fluid
obeys the usual fluid equations of motion with the given effective
equation of state.
In a cosmological context, the effective fluid description (\ref{ee1}) is valid
for the background cosmology and for linear (and, indeed, nonlinear) perturbations. Therefore,
the equation of state of perturbations is the same as that of the
background cosmology, or the so-called entropy perturbation vanishes.

\subsection{Condition for global validity of adiabatic approximation}
\label{domain}
Of course, the adiabatic approximation will not be a good one for all
choices of $V(\phi)$ and $\alpha(\phi)$, or in all physical
situations. Our goal in this subsection is to establish criteria under
which we can trust the adiabatic approximation.
Roughly speaking, the mass of the scalar field
associated with the minimum of the effective potential must be
sufficiently large.
More precisely, we define the effective mass as a function of density
$\rho$ by
\be
m_{\rm eff}^2(\rho) = \left. \frac{\partial^2 V_{\rm eff}}{\partial \phi^2}
  (\phi,\rho) \right|_{\phi = \phi_{\rm m}(\rho)},
\label{meffdef}
\ee
where the derivatives are taken at constant $\rho$, and $\phi_{\rm
  m}(\rho)$ is the value of the scalar field which minimizes the
effective potential at a given density $\rho$.
We assume that $m_{\rm eff}^2$ is positive, otherwise there is no
local minimum of the potential and an adiabatic regime does not arise.
For a perturbation with lengthscale or timescale ${\cal L}$ and density
$\rho$ to be in the adiabatic regime, it is necessary that
\be
{\cal L} \gg m_{\rm eff}^{-1}(\rho).
\label{adiabaticok}
\ee
We call this condition the {\it local adiabatic condition}.\footnote{It has also
been called the Compton condition \cite{Hu:2007nk}, since the RHS of
Eq.\ (\ref{adiabaticok}) is the effective Compton wavelength of the field.}
If the condition is satisfied everywhere in spacetime,
then the adiabatic approximation will be good everywhere.  However if
the condition is satisfied in a local region, it does not necessarily
follow that the adiabatic approximation is valid in that region, due
to non-local effects \cite{Khoury:2003aq,Khoury:2003rn,Hu:2007nk}.  This is
discussed further in Sec.\ \ref{sec:nonlocal} below.

We now derive a slightly more precise version of the local adiabatic criterion
(\ref{adiabaticok}).  For a given density $\rho_0$ we define
$\phi_0 = \phi_{\rm m}(\rho_0)$ and
$\delta \phi = \phi - \phi_0$.
We expand the potential as
\be
V(\phi) = V_0 + V_1 \delta \phi + \frac{1}{2} V_2 \delta \phi^2 +
O(\delta \phi^3) \ ,
\ee
and defining $W(\phi) = e^{\alpha(\phi)}$ we similarly expand
\be
W(\phi) = W_0 + W_1 \delta \phi + \frac{1}{2} W_2 \delta \phi^2 +
O(\delta \phi^3) \ .
\ee
The effective mass (\ref{meffdef}) is then given by
\be
m_{\rm eff}^2 \equiv \frac{\partial^2 V_{\rm eff}}{\partial \phi^2}(\phi_0,\rho_0) = V_2 + \rho_0 W_2 \ ,
\ee
and
the condition (\ref{eq:alg}) that the effective potential be minimized yields
\be
V_1 = -\rho_0 W_1.
\label{ident}
\ee
Also a short computation using the definition (\ref{eq:alg}) of the function $\phi_{\rm m}(\rho)$ gives
\be
\frac{d \phi_{\rm m}}{d\rho} = - \frac{W_1}{V_2 + \rho_0
  W_2} = - \frac{ \alpha_0^\prime e^{\alpha_0}}{m_{\rm eff}(\rho_0)^2} \ ,
\label{derivformula}
\ee
where $\alpha_0 = \alpha(\phi_0)$ and $\alpha_0^\prime = \alpha'(\phi_0)$.
We also define ${\cal L}$ to be the smallest lengthscale or timescale over which the density $\rho$
changes, so that $\nabla_a \nabla^a \rho \sim  \rho/{\cal L}^2$ and
$(\nabla \rho)^2 \sim \rho^2 / {\cal L}^2$.

The field equations are the Einstein equation
(\ref{ee}), the scalar field equation (\ref{eq:scalar1}), and the
fluid equations (\ref{eq:fluid1}) and (\ref{eq:fluid2}).
These equations are not all independent, since the Einstein equation
enforces conservation of the total stress energy tensor.  We will take
as the independent equations just the Einstein equation (\ref{ee}) and
the fluid equations (\ref{eq:fluid1}) and (\ref{eq:fluid2}), since the
scalar field equation can be derived from these.\footnote{Similarly,
the approximate form (\ref{eq:alg}) of the scalar field equation can be
obtained from the approximate form (\ref{ee1}) of the Einstein equation
together with the fluid equations.}

Therefore, to justify the
adiabatic approximation, it is sufficient to justify dropping the scalar
field derivative terms in the Einstein equation (\ref{ee}).
The
ratio of the scalar field gradient terms to the potential term
evaluated at the adiabatic solution is of order
\bea
\frac{ (\nabla \phi)^2}{V(\phi)} \sim
\frac{1}{V_0} \left( \frac{d \phi_{\rm
        m}}{d \rho} \right)^2 ( \nabla \rho )^2
\sim
\frac{1}{V_0 {\cal L}^2} \left( \frac{d \phi_{\rm
        m}}{d \ln \rho} \right)^2.\ \ \
\label{einsteincheck}
\eea
By combining Eqs.\ (\ref{ident}) and (\ref{derivformula}) we obtain
\be
\frac{d \phi_{\rm m}}{d \ln \rho} = \frac{V_1}{m_{\rm eff}^2},
\ee
and using this to eliminate one of the factors of $d \phi_{\rm m} / d
\ln \rho$ from Eq.\ (\ref{einsteincheck}) gives
\bea
\frac{ (\nabla \phi)^2}{V(\phi)} \sim
\frac{V_1}{V_0} \frac{d \phi_{\rm
        m}}{d \ln \rho} \left( \frac{1}{m_{\rm eff}^2 {\cal L}^2}
    \right)
\sim \frac{ d \ln V }{d \ln \rho}
\left( \frac{1} {m_{\rm eff}^2 {\cal L}^2 } \right).\ \ \ \
\label{einsteincheck1}
\eea
Thus, the adiabatic approximation will be valid whenever
\be
\frac{ d \ln V }{d \ln \rho} \left(
\frac{1} {m_{\rm eff}^2 {\cal L}^2 }\right)  \ll 1.
\label{einsteincheck1a}
\ee
Now, since the first factor on the left hand side
involves derivatives of logarithmic
factors, we expect this prefactor to generically be of order unity.
When this is true the condition (\ref{einsteincheck1a}) reduces to the
local adiabatic condition (\ref{adiabaticok}).

\subsection{Nonlocal condition for breakdown of adiabatic approximation}
\label{sec:nonlocal}

As mentioned above, the condition ${\cal L} \gg m_{\rm eff}^{-1}$ in a
local region is a necessary but not a sufficient condition for the validity of
the adiabatic approximation in that region.  This is because
the corrections to the adiabatic approximation are determined by a
wave equation obtained by perturbing Eq.\ (\ref{eq:scalar1}) whose solutions depend in a non-local
way on its sources.  The corrections in a given region can become
large due to a breakdown of the local adiabatic condition (\ref{adiabaticok})
that occurs elsewhere \cite{Hu:2007nk}.

For the special case of static, spherically symmetric
systems, and for a constant density object in a constant density background,
the chameleon field papers \cite{Khoury:2003aq,Khoury:2003rn}
derived a precise condition for the validity of the adiabatic approximation
inside the system, the so-called ``thin shell'' condition, which depends on the asymptotic value
of the potential.  The thin-shell condition is both a necessary and
sufficient condition, but it restricted to static situations.  Below
we will show that for several specific examples
the thin-shell condition and the adiabatic condition
(\ref{adiabaticok}) give the same predictions in order
of magnitude for the boundary of the adiabatic regime.

We now discuss an order-of-magnitude non-local criterion for static,
spherically symmetric situations that predicts when
the adiabatic approximation breaks down even when the local adiabatic condition
(\ref{adiabaticok}) is satisfied.
The condition is a generalization of a condition derived
by Hu \cite{Hu:2007nk} in the context of $f(R)$ modified gravity models, and is
also a generalization of the thin-shell condition of Refs.\ \cite{Khoury:2003aq,Khoury:2003rn}.

Consider a spherically symmetric density profile $\rho(r)$, which we
will assume for simplicity is monotonically decreasing as $r$
increases.  From this density profile we can compute the corresponding
adiabatic scalar field profile
\be
\phi_{\rm ad}(r) \equiv \phi_{\rm m}[\rho(r)].
\label{adprofile}
\ee
If the local adiabatic condition (\ref{adiabaticok}) is
satisfied for all $r$, then this adiabatic field is a good
approximation to the actual solution $\phi(r)$.  Suppose therefore
that the local adiabatic condition is violated\footnote{If the density
  goes to a constant at large $r$, then the local adiabatic condition
  is always satisfied as $r \to \infty$.} for some set of values
${\bar r}_1 < r < {\bar r}_2$.  Then typically what occurs is
that there is a larger interval $r_1 < r < r_2$ with $r_1 < {\bar
  r}_1$ and $r_2 > {\bar r}_2$ on which the solution $\phi(r)$ differs
significantly from the adiabatic field profile (\ref{adprofile}).
For a given interval $(r_1,r_2)$,
significant deviations from the adiabatic approximation should occur in the vicinity of $r=r_1$ if
\be
\phi_2 - \phi_1 \agt \frac{ \alpha_2^\prime e^{\alpha_2}}{r_1} \int_{r_1}^{r_2} r^2 \left[ \rho(r) -
\rho_2\right],
\label{nonlocal}
\ee
where $\rho_i = \rho(r_i)$, $\phi_i = \phi_{\rm ad}(r_i)$, $\alpha_i =
\alpha(\phi_i)$ and $\alpha_i^\prime = \alpha^\prime(\phi_i)$, $i = 1,2$.
The criterion assumes that $r_2 \gg r_1$ and $\rho_2 \ll \rho_1$.

The criterion (\ref{nonlocal}) is a generalization of other criteria that have
appeared in the literature. First, if one assumes that the density
varies on a lengthscale of order $\sim r$, then using this to
approximate the integral in (\ref{nonlocal}) gives
\be
\phi_2 - \phi_1 \agt \alpha_2^\prime e^{\alpha_2} r_1^2 (\rho_1 -
\rho_2).
\label{nonlocal1}
\ee
Equation (\ref{nonlocal1}) is the condition derived in Ref.\ \cite{Hu:2007nk} in the context of
$f(R)$ models, although there the factor of $\alpha_2^\prime
e^{\alpha_2}$ was neglected.  The condition (\ref{nonlocal1}) was
found to reliably predict the onset of deviations from the adiabatic
profile in numerical solutions for the Solar System and the Galaxy\cite{Hu:2007nk}.

Second, one can derive from Eq.\ (\ref{nonlocal}) the
thin-shell condition of Refs.\ \cite{Khoury:2003aq,Khoury:2003rn}, up
to a factor of order unity.
Suppose one has a uniform density sphere of density $\rho_c$ and
radius $R_c$, embedded in a uniform density, infinite medium of
density $\rho_\infty$.  For this case the local adiabatic criterion is
satisfied everywhere except at the point of discontinuity of the
density at $r = R_c$, so we expect a breakdown of the adiabatic
approximation near this point.  Following Ref.\ \cite{Khoury:2003aq}
we assume that $\alpha^\prime = \beta/\mpl$ is a constant, and that we
are in the regime where $\alpha(\phi) \ll 1$, so $e^\alpha \approx
1$.  If we apply the criterion (\ref{nonlocal}) to an interval
$(r_1,r_2)$ with $r_1 < R_c < r_2$, we obtain that
the adiabatic approximation should fail near $r = r_1$
if
\be
\frac{ \Delta \phi} {\beta \mpl \Phi_c} \agt \frac{R_c^3 - r_1^3}{r_1
  R_c^2},
\label{thinshell1}
\ee
where $\Phi_c \sim \rho_c R_c^2/\mpl^2$ is the Newtonian potential at
the center of the sphere.  If the left hand side is large compared to
unity, then Eq.\ (\ref{thinshell1}) will be satisfied for all values
of $r_1$ except very close to $r_1=0$, and it follows that the
adiabatic approximation will not apply in most of the interior of the
sphere.  On the other hand, if the left hand side of Eq.\
(\ref{thinshell1}) is small compared to unity, then it follows that
one would expect the adiabatic approximation to be valid throughout the interior of the
sphere except very close to the surface, in a thin shell of thickness
$\sim R_c \Delta \phi / (\beta \mpl \Phi_c)$.
Both of these conclusions agree in order of magnitude with those of
Refs.\ \cite{Khoury:2003aq,Khoury:2003rn} \footnote{This example shows
that the condition (\ref{nonlocal1}) derived in Ref.\ \cite{Hu:2007nk} is less
general than the condition (\ref{nonlocal}) derived here, since that
condition cannot be used to derive the thickness of the thin shell.}.

We now turn to the derivation of the non-local condition
(\ref{nonlocal}).
We start by noting that static
solutions of the equation of motion (\ref{eq:scalar2}) can be obtained
by extremizing the energy functional
\be
E = \int dr \, r^2 \, \left[ \frac{1}{2} \left( \frac{d \phi}{d r}
  \right)^2 + V(\phi) + e^{\alpha(\phi)} \rho(r) \right].
\ee
The basic idea \cite{Hu:2007nk} is that the adiabatic field profile (\ref{adprofile})
minimizes just the potential energy, and there some kinetic energy
cost for following this adiabatic profile.  When this kinetic energy cost becomes
sufficiently large, it becomes energetically favorable for the field
to switch to a different, non-adiabatic profile with a smaller kinetic
energy and with a larger potential energy, for a net gain in energy.

We now compute the total energies $E_{\rm ad}$ for the adiabatic
profile and $E_{\rm trial}$ for an alternative trial profile which is
qualitatively similar to numerical solutions \cite{Hu:2007nk}.  When
\be
E_{\rm trial} - E_{\rm ad} < 0
\label{condt0}
\ee
there is a net gain in energy for switching to the trial profile, and
so the adiabatic profile is no longer a good approximation.
The trial profile is simply $\phi_{\rm trial}(r) = \phi_2 = $ constant for
$r_1 + \Delta r \le r \le r_2$, and $\phi_{\rm trial}(r) = \alpha - \beta/r$ for
$r_1 \le r \le r + \Delta r$,
with $\alpha$ and $\beta$ chosen to satisfy
continuity at $r_1$ and $r_1 + \Delta r$, $\phi_{\rm trial}(r_1) = \phi_1$ and
$\phi_{\rm trial}(r_1 + \Delta r) = \phi_2$.  This trial profile depends on one
parameter, namely the width $\Delta r$ of the region in which the
field transitions from $\phi_1$ to $\phi_2$.
We obtain
\bea
\label{long}
E_{\rm trial} - E_{\rm ad} &=&
\int_{r_1}^{r_2} dr r^2 \left\{  - \frac{1}{2}  \phi_{\rm ad}'(r)^2   \right. \nn \\
&&   + \alpha_2^\prime e^{\alpha_2} \left[ \phi_{\rm trial}(r) - \phi_{\rm
      ad}(r) \right] \left[ \rho(r) -\rho_2 \right]  \bigg\} \nn \\
&& + \frac{r_1 (r_1 + \Delta r) (\phi_2 - \phi_1)^2 }{2 \Delta r}.
\eea
In deriving this formula we have used an approximate Taylor expansion
of the effective potential about the point $\phi_2$ together with Eq.\ (\ref{eq:alg}):
\bea
V_{\rm eff}(\phi_2,\rho) - V_{\rm eff}(\phi,\rho)
&\approx& ( \phi_2 - \phi) \frac{\partial
  V_{\rm eff}}{\partial \phi}(\phi_2,\rho) \nn \\
&=&  \alpha_2^\prime e^{\alpha_2} ( \phi_2 - \phi) (\rho - \rho_2).\ \ \ \ \ \
\eea

Consider first the kinetic energy contributions, the first and third
lines of Eq.\ (\ref{long}).  The minimum kinetic energy for
any field configuration which satisfies $\phi(r_1) = \phi_1$ and $\phi(r_2) = \phi_2$ is
the energy of the Laplace equation solution with these boundary
conditions\footnote{This can be shown using the Schwarz inequality
  $(\int f g dr)^2 \le \int f^2 dr \, \int g^2 dr$, with $f = r
  \phi'(r)$ and $g = 1/r$.}, namely
$E_{\rm K,min} = (\Delta \phi)^2 r_1 r_2 / (2 (r_2 - r_1))$, where
$\Delta \phi = \phi_2 - \phi_1$.  Therefore we can write the kinetic energy of the
adiabatic field profile [the negative of the first line of Eq.\ (\ref{long})] as
$E_{\rm K,ad} = E_{\rm K,min} (1 + \varepsilon)$, where $\varepsilon$
is dimensionless and nonnegative.
We will assume that $\varepsilon \agt 1$, since $\varepsilon \ll 1$ would require
the adiabatic profile $\phi_{\rm ad}(r)$ to be very close to the $1/r$
profile, which would be a fine tuning.  For generic density profiles
we expect $\varepsilon \agt 1$.
If we now choose $\Delta r = 2 r_1 /
\varepsilon$, then it follows that the net gain in kinetic energy is
\be
\varepsilon \frac{E_{\rm K,min}}{2} \agt E_{\rm K,{\rm min}} \sim r_1 \Delta
\phi^2,
\label{KEgain}
\ee
using $r_2 \gg r_1$.

Turn now to the potential energy contributions, the second line of
Eq.\ (\ref{long}).  If we assume that the potential $V(\phi)$ is
monotonically decreasing, as in the Chameleon field models, then it
follows that $\phi_{\rm ad}(r)$ is an increasing function of $r$, so
that
\be
\phi_{\rm trial}(r) - \phi_{\rm ad}(r) \le \phi_2 - \phi_1.
\ee
This implies that the potential energy term in Eq.\ (\ref{long}) is
bounded above by
$$
 \alpha_2^\prime e^{\alpha_2} \Delta \phi \int_{r_1}^{r_2} r^2 \left[ \rho(r) -
\rho_2\right].
$$
Inserting this together with the estimate (\ref{KEgain}) of the kinetic
energy gain into Eq.\ (\ref{condt0}) yields the criterion (\ref{nonlocal}).

\subsection{Observed equation of state parameter}

In this section we derive the observed equation of state parameter
$w_{\rm obs}$ for these models, in the adiabatic regime, for the case $\alpha_b =0$.
Generically we have $w_{\rm
  obs} < -1$ corresponding to superacceleration, as previously noted
in Ref.\ \cite{Das:2005yj}.

The background cosmological evolution is given in the Einstein frame
by the equation
\be
3 \mpl^2 H^2 = V + e^{\alpha} \frac{\rho_0}{a^3},
\label{eee1}
\ee
where $\rho_0$ is a constant, together with the evolution equation for
the scalar field.  Observations of the acceleration of the Universe
are fit to the model
\be
3 \mpl^2 H^2 = \frac{\rho_1}{a^3} + \rho_{\rm DE}(a)
\label{eee2}
\ee
where $\rho_1 = e^{\alpha_0} \rho_0$ is the observed matter density
today\footnote{Note that in this model some fraction of the mass
  density in the $V$ term in Eq.\ (\ref{eee1}) will cluster, so
  measurements of mass density using clusters will not correspond
  exactly to measurements of the second term in Eq.\ (\ref{eee1}).  We
  will not address this issue here.}, $\alpha_0$ is the value of $\alpha$ today,
and $\rho_{\rm DE}$
is the inferred ``dark energy density''.
The equation of state parameter is then given by
\be
w_{\rm obs}(a) = -1 - \frac{1}{3} \frac{d \ln \rho_{\rm DE}}{d \ln a}.
\label{eee3}
\ee
Combining Eqs.\ (\ref{eee1}) -- (\ref{eee3}) and using Eq.\ (\ref{eq:alg})
to eliminate the term proportional to $d\phi/da$ gives
\be
w_{\rm obs} = \frac{- V}{ V + \frac{e^{\alpha_0} \rho_0}{a^3}
  (e^{\alpha - \alpha_0} -1)}.
\ee
Next we use Eq.\ (\ref{eq:alg}) again to obtain $\rho_0/a^3 = -
e^{-\alpha} V'(\phi) / \alpha'(\phi)$, which gives
\be
w_{\rm obs} = \frac{-1}{1 - \frac{d \ln V}{d \alpha} ( 1 - e^{\alpha_0
    - \alpha})}.
\ee

This formula has the property that $w_{\rm obs}=-1$ today for all
models. Also expanding to first order about $a=1$ we obtain $1/w_{\rm
  obs} = -1 + \ln(V/V_0)$, where $V_0$ is the value of $V$ today, so
$w_{\rm obs} < -1$ in the past since $V = V[\phi_{\rm m}(\rho)]> V_0$ in the past.

\section{Adiabatic instability}
\label{instability}

In the adiabatic regime, the models (\ref{action0}) discussed here can
exhibit instabilities on
small scales characterized by a negative sound speed squared of the
effective coupled fluid.
This instability extends down to the
smallest scales for which the adiabatic approximation is valid.
Starting with a uniform fluid, the instability will give rise to
exponential growth of small perturbations.  The final state of the
coupled fluid is beyond the scope of this paper.
Theories that exhibit this instability are typically ruled out as
models of dark energy.

\subsection{Hydrodynamic viewpoint}
\label{sec:hydro}

This adiabatic instability was first discovered by Afshordi,
Zaldarriaga and Kohri \cite{Afshordi:2005ym} in a context slightly
different to that considered here.  That context was the mass varying neutrino model of
dark energy, where a dynamical dark energy model is obtained by
coupling a light scalar field to neutrinos but not to dark matter.
The instability was previously discussed in the context of the models
considered here by Kaplinghat and Rajaraman \cite{Kaplinghat:2006jk}.
In this paper we will generalize
the treatment of the instability given in Ref.\ \cite{Kaplinghat:2006jk}.

A fairly simple form of the stability criterion can be obtained by
writing the potential $V(\phi)$ as a function $V(\alpha)$ of the
coupling function $\alpha(\phi)$ by eliminating $\phi$.
This gives
\be
\rho_{\rm eff} = V + e^{ \alpha} \rho = V - \frac{
  dV/d\phi}{d\alpha/d\phi} = V - \frac{dV}{d\alpha} \ ,
\ee
where we have used Eqs.\ ~(\ref{rhoeff}) and (\ref{eq:alg}).  The
square $c_a^2$ of the adiabatic sound speed is then given by
\be
\frac{1}{c_a^2} = \frac{d \rho_{\rm eff}}{dp_{\rm eff}} = \frac{
  d\rho_{\rm eff}/d\alpha} {dp_{\rm eff}/d\alpha} = \frac
{\frac{d}{d\alpha} \left[ V - \frac{dV}{d\alpha} \right] }
{\frac{d}{d\alpha} \left[ -V \right] } = -1 + \frac{ \frac{d^2 V}{d\alpha^2} }{
  \frac{dV}{d\alpha}} \ .
\label{soundspeed}
\ee
The system will be unstable if
\be
c_a^2 <0.
\label{unstablec}
\ee
Equation (\ref{soundspeed}) gives a simple prescription for computing when a
given theory will be unstable, assuming it is in the adiabatic
regime.  This equation furnishes $c_a^2$ as a function of $\alpha$,
which can be re-expressed as a function of $\phi$ using $\alpha =
\alpha(\phi)$, and then as a function of density $\rho$ using $\phi =
\phi_{\rm m}(\rho)$ from Eq.\ (\ref{eq:alg}).

More generally, outside of the adiabatic regime, the
effective sound speed is the ratio of the local pressure and density
perturbations
$$
c_s^2 = \frac{\delta p}{\delta \rho}
$$
and can differ from $c_a^2$.  In a cosmological context we have
$c_a^2\equiv \dot{P}/\dot{\rho}$ and $c_s^2(k,a)\equiv \delta
P(k,a)/\delta\rho(k,a)$, where $k$ is spatial wavenumber and $a$ is
scale factor.  For most of this paper we will consider only the
adiabatic regime in which $c_s^2 \to c_a^2$, although the more general
regime will be probed in our numerical integrations in Sec.\ \ref{examples}.

We now argue that the adiabatic sound speed (\ref{soundspeed}) is {\it always}
negative in the adiabatic regime.
In the definition (\ref{meffdef}) of the effective mass, we rewrite
the $\phi$ derivative in terms of $\alpha$ derivatives using
$\alpha = \alpha(\phi)$.
Using the fact that the first derivative of $V_{\rm eff}$ with respect to $\phi$ vanishes at
$\phi = \phi_{\rm m}(\rho)$, we obtain
\be
m_{\rm eff}^2 = \left( \frac{ d\alpha }{d \phi} \right)^2
\frac{\partial ^2 V_{\rm eff}}{\partial \alpha^2},
\ee
where the derivatives are taken at constant $\rho$.
Using Eq.\ (\ref{phieqn}) and then eliminating $\rho$ using Eq.\
(\ref{eq:alg})
gives
\be
m_{\rm eff}^2 = \left( \frac{ d\alpha }{d \phi} \right)^2
\left[ \frac{d^2 V}{d\alpha^2} - \frac{d V}{d\alpha} \right].
\label{meff2}
\ee
Simplifying using the expression (\ref{soundspeed}) for the sound speed squared
gives
\be
m_{\rm eff}^2 = \left( \frac{ d\alpha }{d \phi} \right)^2
 \frac{d V}{d\alpha} \frac{1}{c_s^2}.
\label{meff3}
\ee
The adiabatic instability arises in the
case $c_s^2<0$ and $m_{\rm eff}^2>0$. [When $m_{\rm eff}^2<0$ there is
no adiabatic regime, in the sense that adiabatic solutions are subject
to a tachyonic instability.  This instability
has been discussed in the context of $f(R)$ gravity models in Refs.\
\cite{Dolgov:2003px,Seifert:2007fr,Sawicki:2007tf}.]  It follows from Eqs.\
(\ref{soundspeed}) and (\ref{meff3}) that the adiabatic instability
occurs when
\be
\frac{dV}{d\alpha} < \frac{d^2 V}{d\alpha^2} < 0,
\label{cc1}
\ee
or when
\be
\frac{dV}{d\alpha} < 0 < \frac{d^2 V}{d\alpha^2}.
\label{cc2}
\ee

However, the region in the $(V_{,\alpha},V_{,\alpha\alpha})$ parameter
space defined by the union of the regions (\ref{cc1}) and (\ref{cc2}) is
just the adiabatic regime; the adiabatic approximation requires
both that $V_{,\alpha}<0$ so that the effective potential has a local
extremum, and also that $m_{\rm eff}^2>0$ so that the local extremum is a
local minimum.  It follows that the adiabatic instability condition
(\ref{unstablec}) is satisfied throughout the adiabatic regime.
This genericity of the instability has also been deduced by Kaplinghat
and Rajaraman \cite{Kaplinghat:2006jk} using a different method.
However, as we discuss in the following subsection, the instability
actually occurs only if the coupling $\alpha'(\phi)$ is sufficiently
large, a point missed in Refs.\ \cite{Afshordi:2005ym,Kaplinghat:2006jk}.

\subsection{Scales over which the instability operates}
\label{sec:scales}

For the instability to be relevant on some spatial scale ${\cal L}$, then we must have
\be
{\cal L} \gg m_{\rm eff}^{-1}
\ee
in order that the adiabatic approximation be valid, as discussed
above.  There is also an upper bound on the range of spatial scales
which comes about as follows.  When the instability is present,
spatial Fourier modes with wavelength ${\cal L}$ grow exponentially on
a timescale
\be
\tau \sim \frac{{\cal L}}{\sqrt{|c_s^2|}} \ ,
\ee
where $c_s^2$ is given by Eq.~(\ref{soundspeed}).  If this timescale
is longer than the Hubble time $H^{-1}$, then the mode does
not have time to grow and the instability is not relevant.
Therefore the range of scales over which the instability operates is
\be
m_{\rm eff}^{-1} \ll {\cal L} \ll \frac{\sqrt{|c_s^2|}}{H} \ .
\ee
More generally,
for a fluid of density $\rho$, if the instability is to be unmodified
by the gravitational dynamics of the fluid, then the instability
timescale must be shorter than the gravitational dynamical time, which
from Eq.\ (\ref{ee1}) is $\sim \mpl/\sqrt{\rho_{\rm eff}(\rho)}$.
Here $\rho_{\rm eff}(\rho)$ is the total mass density (\ref{rhoeff}) of the
coupled dark matter-dark energy fluid.
This gives the criterion
\be
m_{\rm eff}(\rho)^{-1} \ll {\cal L} \ll \frac{\mpl
  \sqrt{|c_s^2(\rho)|}}{\sqrt{\rho_{\rm eff}(\rho)}} \ ,
\label{range}
\ee
which determines the values of density $\rho$ and lengthscale ${\cal
  L}$ for which the instability operates.
The upper lengthscale can be rewritten using Eqs.\ (\ref{meff3}),
(\ref{rhoeff}) and (\ref{eq:alg}) to give
\be
\frac{1}{m_{\rm eff}(\rho)} \ll  {\cal L} \ll \frac{\mpl
|\alpha^\prime[\phi_{\rm m}(\rho)]|}{m_{\rm eff}(\rho)}
\sqrt{\frac{1}{1 - \frac{1}{\frac{d\ln V}{d \alpha}}}}.
\label{range1}
\ee
Here the quantity $d \ln V / d\alpha(\alpha)$
on the right hand side is
expressed as a function of $\phi$ using $\alpha = \alpha(\phi)$, and
then as a function of the density using $\phi = \phi_{\rm m}(\rho)$.

At longer lengthscales, it is possible that the negative sound speed squared still
engenders an instability, but determining this requires a stability
analysis of the system in question including the effects of self
gravity. The results might vary from one system
to another.
In this paper we will restrict attention to the regime (\ref{range}) where
the presence of the instability can be easily diagnosed.

The factor in square brackets in Eq.\ (\ref{range1}) is always smaller
than unity, since $V$ is assumed to be a decreasing function of $\phi$
and hence also of $\alpha$.  It follows that the ratio of the maximum
lengthscale ${\cal L}_{\rm max}$ to the minimum lengthscale ${\cal
  L}_{\rm min}$ satisfies
\be
\frac{{\cal L}_{\rm max}}{{\cal L}_{\rm min}} \le \mpl
|\alpha'[\phi_{\rm m}(\rho)]|.
\ee
Hence in order for there to be a nonempty regime in which the
instability operates, the strength of the coupling of the field to the
dark matter must be much stronger than gravitational strength,
\be
\mpl |\alpha' |\gg 1,
\label{strongcoupling}
\ee
as discussed in the introduction.  In Sec.\ \ref{jeans} below we give a simple
explanation for this requirement.
In Sec.\ \ref{sec:MaVaN} below we give an example of a model where we confirm
numerically instability is present at strong coupling but not when the
coupling is weak.

\subsection{Jeans instability viewpoint}
\label{jeans}

There are two different ways of describing and understanding the instability, depending
on whether one thinks of the
scalar-field mediated forces as being ``gravitational'' forces or ``pressure'' forces.
From one point of view, that of the Einstein frame description, the
instability is independent of gravity.  This can be seen from the
equation of motion (\ref{ee1}); the instability is present even when
the (Einstein-frame) metric perturbation due to the fluid can be
neglected.  In the adiabatic regime the acceleration due the scalar
field is a gradient of a local function of the density [cf.\ Eq.\
(\ref{euler1}) below], which can be thought of as a pressure.  The net
effect of the scalar interaction is to give a contribution to the
specific enthalpy $h(\rho) = \int dp/\rho$ of any fluid which is independent
of the composition of the fluid.
If net sound speed squared of the
fluid is negative, then there exists an instability in accord with our
usual hydrodynamic intuition.

From another point of view, however, that of the Jordan
frame description, the instability involves gravity.
The gravitational force in this frame is mediated partly by a tensor
interaction and partly by a scalar interaction.
The effective Newton's constant describing the interaction of dark matter with itself is
\be
G_{cc} = G \left[ 1
  + \frac{2 \mpl^2 \alpha^\prime(\phi)^2 }{1 +
\frac{m_{\rm eff}^2}{ {\bf k}^2 }
} \right],
\label{Gformula0}
\ee
where ${\bf k}$ is a spatial wavevector.
This is Eq.\ (\ref{Gformula}) of Appendix \ref{sec:Geff} specialized to
${\rm i} = {\rm j} = c$, with $\alpha_c$ written just as $\alpha$, and
specialized to $\alpha_b=0$ (since
experiments tell us that $|\alpha_b'| \alt 10^{-2} \mpl^{-1}$ today).
Here the 1 in the square brackets describes the tensor interaction and
the second term the scalar interaction.
At long lengthscales, $k \ll m_{\rm eff}/(\mpl |\alpha'|)$, the scalar interaction is
suppressed and we have $G_{cc} \approx G$.  At short lengthscales,
$k \gg m_{\rm eff}$, the scalar field is effectively massless and
$G_{cc}$ asymptotes to a constant, $G_{cc} \approx G [1 + 2 \mpl^2 (\alpha^\prime)^2 ]$.
However, when $\mpl |\alpha^\prime| \gg 1$ there is an intermediate range of
lengthscales,
\be
\frac{m_{\rm eff}} {\mpl |\alpha^\prime|} \ll k \ll m_{\rm eff}
\label{range3}
\ee
in which the effective Newton's constant increases linearly with
$k^2$,
\be
G_{cc} \approx G \frac{2 \mpl^2 (\alpha^\prime)^2}{m_{\rm eff}^2}
{\bf k}^2.
\label{Gcc1}
\ee
A gravitational interaction with $G_{cc} \propto {\bf k}^2$ behaves just
like a (negative) pressure in the hydrodynamic equations.  This
explains why the the effect of the scalar interaction can be thought
of as either pressure or gravity in the range of scales (\ref{range3}).
Note that the range of scales (\ref{range3}) coincides with with the
range (\ref{range1}) derived above, up to a logarithmic correction factor.

From this second, Jordan-frame point of view, the instability is
simply a Jeans instability.  In a cosmological background with
Einstein-frame metric $ds^2 = -dt^2 + a(t)^2 d{\bf x}^2$, the evolution equation for
the CDM fractional density perturbation $\delta$ with comoving
wavenumber $k_c$ on subhorizon scales in the adiabatic limit is \cite{Brax:2005ew,paperIII}
\be
{\ddot \delta} + 2 H {\dot \delta} - 4 \pi G_{\rm cc} e^\alpha \rho
\delta = 0,
\label{perts}
\ee
where $H = {\dot a}/a$.  Here $G_{cc}$ is given by the expression
(\ref{Gformula0}) evaluated at the physical wavenumber $k= k_c/a$,
and we have neglected photons and baryons.

Now in the absence of the Hubble damping term in Eq.\ (\ref{perts}),
the gravitational
interaction described by the last term would cause an exponential
growth of the mode, the usual Jeans instability of uniform fluid.
Normally in a cosmological context, the Hubble damping term
is present and the timescale $\sim 1/H$ associated with this term is
of the same order as the timescale $1/ \sqrt{G \rho}$ associated with
the gravitational interaction in the last term.  Because of this
equality of timescales, the exponential growth is converted to power
law growth by the Hubble damping.  In the present context, however,
things work differently.  The gravitational constant governing the
gravitational self-interaction of the mode is $G_{cc}(k)$ instead of
$G$, and consequently the timescale associated with the last term in
Eq. (\ref{perts}) is shorter than the Hubble damping time by
a factor of
\be
\sim \sqrt{\frac{G_{cc} }{ G }} \sim \frac{ k \mpl |\alpha^{\prime}|  }{ m_{\rm
  eff}} \gg 1.
\ee
Therefore the Hubble damping is ineffective and the Jeans instability
causes approximate exponential growth rather than power law growth.

The above discussion can also be cast in terms a scale-dependent sound speed instead of a scale
dependent Newton's constant; this clarifies the relation to our previous
discussion of Secs.\ \ref{sec:hydro} and \ref{sec:scales}.
As a slight generalization, consider
a fluid with an intrinsic sound speed $c_{\rm s,in}$.  Then
the evolution equation (\ref{perts}) generalizes to (see Appendix \ref{visible})
\be
{\ddot \delta} + 2 H {\dot \delta} + \frac{ c_{\rm tot}(k)^2 k^2 }{a^2} \delta= 0,
\label{pertsa}
\ee
where the effective total sound speed squared is
\be
c_{\rm tot}(k)^2 = c_{\rm s,in}^2 - \frac{4 \pi}{k^2} G e^\alpha \rho
\left[ 1
  + \frac{2 \mpl^2 \alpha^\prime(\phi)^2 }{1 +
\frac{m_{\rm eff}^2}{ {\bf k}^2 }
} \right].
\label{ctot}
\ee
In the range of lengthscales (\ref{range3}) this squared sound speed is a
constant, independent of $k$, as for a normal, hydrodynamic sound
speed.  Outside of this range of scales, we have $c_{\rm tot}^2 \propto
1/k$ at both large and small $k$ (if the intrinsic sound speed can be neglected), describing a conventional
gravitational interaction.

We reiterate that the existence of the range of scales (\ref{range3})
in which Newton's constant scales linearly with $k^2$ depends on the
assumption of strong coupling, $|\alpha'| \mpl \ll 1$.  If, instead,
$|\alpha'| \mpl \alt 1$, the dependence of $G$ on $k$ is very close to
that of standard gravity, and the instability reduces to
the normal Jeans instability of a fluid, characterized in a
cosmological context by power law growth.

\subsection{Domain of validity of fluid description of dark matter}

Up till now we have described cold dark matter as a pressureless
fluid.  However, at a more fundamental level, one should use a kinetic
theory description based on the collisionless Boltzmann equation.  In
the conventional $\Lambda$CDM framework, the fluid approximation
breaks down at small scales, below the free-streaming lengthscale,
and also in the nonlinear regime after violent relaxation has taken
place in CDM halos \cite{Gunn}.  We now discuss how, in the models discussed here,
the conventional picture for the fluid domain of validity is slightly modified.

Let us denote by $\sigma$ the rms velocity of the dark matter particles.
Consider a perturbation characterized by a wavelength $\lambda$ and
wavenumber $k = 2 \pi / \lambda$.
The characteristic growth or oscillation time associated with this perturbation is
$
\tau(k) \sim \lambda / \sqrt{ | c_{\rm tot}(k) |^2},
$
where the total effective sound speed $c_{\rm tot}$ is given by Eq.\ (\ref{ctot}).
The distance traveled by a dark matter particle in this time is $d(k)
\sim \sigma \tau(k)$, and the ratio of this distance to the size of
the perturbation is
\be
\frac{d(k) }{\lambda} \sim \frac{\sigma \tau(k)}{\lambda} \sim
\frac{\sigma}{\sqrt{|c_{\rm tot}(k)^2|}}.
\label{ratio0}
\ee
When this dimensionless ratio is of order unity or larger,
perturbations do not have time to grow before they are washed out by
free streaming of the particles, and the fluid approximation breaks
down.  Using the formula (\ref{ctot}) with $c_{\rm s,in}$ set to
zero\footnote{Since we expect $c_{\rm s,in} \sim \sigma$, setting
$c_{\rm s,in}$ to zero is only consistent in the regime $\sigma^2
\ll | c_s^2|$.  However, for $\sigma^2 \agt |c_s^2|$, retaining the intrinsic sound speed in Eq.\
(\protect{\ref{ratio0}}) does not change the final result
(\protect{\ref{lambdaFS}}) for the free streaming scale in order of
magnitude.},
, we obtain
\be
\frac{d(k)}{\lambda} \sim \frac{ \sigma k}{ \sqrt{4 \pi G e^\alpha \rho}}
\left[ 1
  + \frac{2 \mpl^2 \alpha^\prime(\phi)^2 }{1 +
\frac{m_{\rm eff}^2}{ {\bf k}^2 }
} \right]^{-1/2}.
\label{ratio1}
\ee

Now in the conventional CDM framework, the factor in the square
brackets is unity, so the ratio (\ref{ratio1}) is proportional
to $k$ and becomes large as $k\to\infty$.  Hence the fluid approximation breaks down on small scales,
below the critical free-streaming lengthscale $\lambda_{\rm FS} \sim
\sigma / \sqrt{G \rho}$.

In the present context things work a little differently due to the
scale dependence of Newton's constant.
We can rewrite Eq.\ (\ref{ratio1}) in the approximate form
\begin{equation}
\label{ratio2}
\frac{d(k)}{\lambda} \sim \left\{ \begin{array}{ll}
    \frac{\sigma}{\sqrt{|c_s^2|}} \frac{\sqrt{2} k \mpl | \alpha'|}{m_{\rm eff}} & \mbox{ $k \ll
        \frac{m_{\rm eff}}{\sqrt{2} \mpl |\alpha'|}$,}\\
       \frac{\sigma}{\sqrt{|c_s^2|}} & \mbox{
        $\frac{m_{\rm eff}}{\sqrt{2} \mpl |\alpha'|} \ll k \ll m_{\rm
          eff},$}\\
         \frac{\sigma}{\sqrt{|c_s^2|}} \frac{k}{m_{\rm eff}} &
        \mbox{ $ m_{\rm eff} \ll k, $}\\
        \end{array} \right.
\end{equation}
where $c_s^2 = - 8 \pi G e^\alpha \rho \mpl^2 \alpha^{\prime\,2} / m_{\rm eff}^2$
is the constant value of the second term in the
expression (\ref{ctot}) for $c_{\rm tot}^2$ in the range of scales
(\ref{range3}), or equivalently the sound speed discussed in Sec.\ \ref{sec:hydro}.
We see that the ratio $d(k)/\lambda$ is proportional to $k$ at large
scales and at small scales, but that in the intermediate range of
scales it is a constant, so that the effect of free streaming is
equally important for all the modes in this range.
If we define the free streaming lengthscale $\lambda_{\rm FS}(\sigma)$
to be the smallest lengthscale for which free streaming is unimportant, $d(k)/\lambda \alt 1$, then we
obtain
\begin{equation}
\label{lambdaFS}
\lambda_{\rm FS}(\sigma) \sim \left\{ \begin{array}{ll} \frac{\sigma}{\sqrt{|c_s^2|}}
    \frac{ \mpl | \alpha'|}{m_{\rm eff}} & \mbox{ $\sigma^2 \agt
        |c_s^2|$,}\\
       \frac{\sigma}{\sqrt{|c_s^2|}} \frac{1}{m_{\rm eff}}& \mbox{
        $\sigma^2 \alt |c_s^2|.$}\\
        \end{array} \right.
\end{equation}
This lengthscale jumps discontinuously at $\sigma^2 \sim |c_s^2|$.

There are thus two different regimes that occur:
\begin{itemize}

\item When $\sigma^2
\ll |c_s^2|$, free streaming is important only at
scales small compared to $1/m_{\rm eff}$ (for which the adiabatic
approximation is invalid anyway).
The fluid approximation is valid throughout the range of
lengthscales (\ref{range3}), and so the adiabatic instability is present.
This conclusion is confirmed by a kinetic theory analysis (see
Appendix \ref{kinetic}), which shows
that linearized perturbations of any
homogeneous, isotropic initial particle distribution function are always unstable on scales
that are in the regime (\ref{range3}), as long as $\sigma^2 \ll |c_s^2|$.

\item When $\sigma^2 \agt |c_s^2|$, free streaming becomes important
and the fluid approximation breaks down throughout the range of scales
(\ref{range3}).
One expects the free streaming
(also called Landau damping) to kill the instability.
This is confirmed by our kinetic theory analysis of Appendix \ref{kinetic}:
we show that for a Maxwellian distribution, the finite velocity
dispersion stabilizes the coupled fluid whenever
$\sigma^2 \ge | c_s^2|$, in agreement with the analysis of Ref.\ \cite{Afshordi:2005ym}.

\end{itemize}

Consider now the evolution of cosmological perturbation modes.  When
will our analysis of instability apply?  First,
the condition $\sigma^2 \le |c_s^2|$ is not
very restrictive, since the CDM cools rapidly with the Universe's expansion, with
temperature scaling as $(1+z)^2$ once the
particles become non-relativistic.  For example, for CDM particles of
mass $\sim $ GeV with weak scale cross sections, the CDM temperature
is $\sim 10^{-4}$ K at decoupling \cite{Gunn}.
However, after perturbations go nonlinear and
violent relaxation takes place in CDM halos, the effective
coarse-grained velocity dispersion becomes much larger.
Hence, our analysis does not apply to modes that are in the nonlinear
regime.  Our analysis will apply in the early Universe, before any
modes have gone nonlinear.  It will also apply to large scale modes, even
after smaller scale modes have gone nonlinear, since such large scale
modes should still be well described by linear theory (see for example
the qualitative arguments in chapter 28 of Peebles \cite{Peebles}).
Our investigation of specific models later in the paper will focus on
these large scale, linear regime modes.

We note that earlier investigations of the instability focused instead
on small scale modes, below the free streaming scale $\lambda_{\rm
  FS}$ \cite{Afshordi:2005ym,Kaplinghat:2006jk}.
From the formula (\ref{lambdaFS}) we see that, for the models
discussed here, either the adiabatic approximation is not valid on these
small scales since $\lambda_{\rm FS} \alt 1/m_{\rm eff}$, or
$\sigma^2 \agt |c_s^2|$ and the instability is killed by free streaming.

\section{Examples of theories with adiabatic instability}
\label{examples}
In this section we discuss some specific classes of theories.

\subsection{Exponential potential and constant coupling}
\label{sec:exponential}

We first consider theories with exponential potentials of the form
\be
V = V_0 e^{-\lambda \phi/\mpl} \ ,
\label{expV}
\ee
with $\lambda>0$ and with linear coupling functions
\be
\alpha(\phi) = - \beta C \frac{\phi}{\mpl} \ ,
\label{constantcoupling}
\ee
where $\beta = \sqrt{2/3}$ and $C$ is a constant.\footnote{The
  notation in Eq.\ (\ref{constantcoupling}) is chosen such that $f(R)$
  gravity theories correspond to $C=1/2$.}
These theories have been previously studied in Ref.\ \cite{Amendola:1999er}.
The effective potential is, from Eq.\ (\ref{phieqn}),
\be
V_{\rm eff}(\phi,\rho) = V_0 e^{-\lambda \phi/\mpl} + e^{-\beta C \phi/\mpl} \rho \ ,
\ee
and solving for the local minimum of this potential yields
the relation between $\phi$ and $\rho$ in the adiabatic regime:
\be
e^{(\lambda -\beta C) \phi_{\rm m}(\rho)/\mpl} = \frac{ \lambda V_0 }{-\beta C
  \rho} \ .
\label{phimans}
\ee
Note that $C$ must be negative in order for
the effective potential to have a local minimum and for an adiabatic
regime to exist.  We will restrict attention to this case, and we
define the dimensionless
positive parameter $\gamma = - \lambda / \beta C$.
The corresponding effective mass parameter is
\be
m_{\rm eff}^2 = \lambda^2 \mpl^{-2} V_0 \frac{1+\gamma}{\gamma} \left( \frac{
    \rho}{\gamma V_0} \right)^{\frac{\gamma}{\gamma+1}} \ .
\label{mass1}
\ee

Next, we compute the sound speed squared.  Using Eq.\
(\ref{soundspeed}) we obtain
\be
c_s^2  =  - \frac{1}{1+\gamma} \ ,
\label{cs21}
\ee
so this
model is always unstable in the adiabatic regime.
From Eqs.\ (\ref{expV}) and (\ref{phimans}) we also obtain
\be
\frac{\partial \ln V}{\partial \ln \rho} = \frac{\gamma}{1 + \gamma} \ .
\label{logd}
\ee
We now insert the effective mass (\ref{mass1}), the sound speed
squared (\ref{cs21}) and the logarithmic derivative
(\ref{logd})
into Eq.\ (\ref{einsteincheck1a})
and into the second half of
Eq.\ (\ref{range}).  This yields the range of spatial scales ${\cal
  L}$ over which
the instability operates for a given density $\rho$ to be
${\cal L}_{\rm min}(\rho) \ll {\cal L} \ll {\cal L}_{\rm max}(\rho)$,
where
\be
{\cal L}_{\rm min}(\rho)^2 = \frac{\gamma^2 }{\lambda^2 (1 + \gamma)^2 }
\frac{\mpl^2}{V_0}
\left( \frac{ \gamma
    V_0}{ \rho} \right)^{\frac{\gamma}{\gamma+1}}
\label{Lmin2}
\ee
and ${\cal L}_{\rm max}(\rho)^2$ is a constant times this:
\be
{\cal L}_{\rm max}(\rho)^2 = \beta^2 C^2 {\cal L}_{\rm min}(\rho)^2 \ .
\label{Lmax2}
\ee
Thus, there is a nonempty unstable regime only when $\beta |C| \gg 1$,
ie with the scalar coupling is strong compared to the gravitational
coupling, in agreement with the discussion in Sec.\ \ref{sec:scales}.

To see the effect of the instability more explicitly, we consider cosmological perturbations.
The Einstein-frame FRW equation in the adiabatic limit is
\be
3 \mpl^2 H^2 = V + e^\alpha \rho \ ,
\label{cosmo1}
\ee
where $\rho \propto 1/a^3$.  This yields $a(t) \propto t^{2/(3 + 3 w_{\rm eff})}$, where
the effective equation of state parameter is
\be
w_{\rm eff} = - \frac{1}{1 + \gamma} \ .
\label{weff1}
\ee
In the strong coupling limit $|C| \to \infty$ that we specialize to here, $w_{\rm eff} \to -1$.
Thus the adiabatic regime of this model with large $|C|$ is incompatible with
observations in the matter dominated era, where we know $w_{\rm eff} \approx
0$ except for at small redshifts.
Nevertheless, the model is still
useful as an illustration of the instability.

We find from Eqs.\ (\ref{Lmin2}), (\ref{Lmax2}) and (\ref{cosmo1})
that the range of unstable scales is given by
\be
\frac{1}{\beta^2 C^2} \ll \frac{H^2 a^2}{k^2}  \ll \frac{1}{3(1+\gamma)} \ ,
\ee
where $k$ is comoving wavenumber.  This range of scales always lies
just inside the horizon.  A given mode $k$ will evolve through this
unstable region before it exits the horizon.

Next we use the approximate form (\ref{Gcc1}) of Newton's constant in the
perturbation evolution equation (\ref{perts}), and transform from $t$
derivatives to $a$ derivatives.
This gives
\be
\frac{d^2 \delta}{d a^2} + \frac{3}{a} \left( 1 - \frac{1}{2} \frac{d
    \ln \rho_{\rm eff}}{d \ln \rho} \right) \frac{d \delta}{da} -
\left( \frac{ (\alpha^\prime)^2 k^2 e^\alpha \rho}{m_{\rm eff}^2 H^2
    a^4} \right) \delta =0 \ .
\ee
Specializing this equation to the exponential model using Eqs.\
(\ref{constantcoupling}), (\ref{phimans}), (\ref{mass1}) and
(\ref{cosmo1}) and taking the strong coupling limit
$|C| \to \infty$ gives
\be
\frac{d^2 \delta}{d a^2} + \frac{3}{a} \frac{d \delta}{da} -
\frac{k^2}{ H^2 a^4} \delta =0 \ .
\ee
In the strong coupling limit $H$ is approximately a constant, $H
\approx H_0$, and the growing mode solution is
\be
\delta(a) \propto \frac{1}{a} K_1\left(\frac{k}{H_0 a}\right) \approx
\sqrt{ \frac{\pi H_0}{2 k a}} \exp \left(-\frac{k}{H_0 a} \right) \ ,
\ee
where $K_1$ is the modified Bessel function.  The mode grows by a
factor $\sim e$ when the scale factor changes from $a$ to $a + \Delta
a$, where $\Delta a/a \sim a H_0 / k \ll 1$ for subhorizon
modes.

A more detailed analysis of the cosmology of this model is given in
Ref.\ \cite{paperIII}, but in the non-adiabatic regime $|C| \sim 1$ rather than the
strong coupling regime $|C| \gg 1$ considered here.

\subsection{Two component dark matter models}
\label{sec:twocomponent}

We next consider models in which there are two dark matter sectors, a
density $\rho_c$ which is not coupled to the scalar field, and a
density $\rho_{co}$ which is coupled with coupling function (\ref{constantcoupling})
and exponential potential (\ref{expV}).
Both of these components are treated as pressureless fluids.
The FRW equation for this model in the adiabatic limit is
[cf.\ Eq.\ (\ref{cosmo1}) above]
\be
3 \mpl^2 H^2 = V + e^\alpha \rho_{co} + \rho_c \ .
\label{cosmo2}
\ee
Similar two component models have been considered by Farrar and Peebles
\cite{Farrar:2003uw}.  This model is also similar to the mass varying
neutrino model model \cite{Afshordi:2005ym,Bjaelde:2007ki}
where the neutrinos play the role of the coupled component; see Sec.\
\ref{sec:MaVaN} below.
The first two terms on the right hand
side of Eq.\ (\ref{cosmo2}) act like a fluid with equation of state
parameter given by (\ref{weff1}), and in the strong coupling limit
$|C| \gg 1$ this
fluid acts like a cosmological constant.  Thus, the background
cosmology can be made close to $\Lambda$CDM by taking $|C|$ to be
large.

The fraction of dark matter which is coupled must be small in the
limit of large coupling, $|C| \gg 1$.
Denoting $\Omega_V = V / (3 \mpl^2 H^2)$,
$\Omega_{co} = e^\alpha \rho_{co} / (3 \mpl^2 H^2)$ and
$\Omega_c = \rho_c / (3 \mpl^2 H^2)$, we have $1 = \Omega_V +
\Omega_{co} + \Omega_c$.  Also from Eq.\ (\ref{phimans}) it follows
that, if the asymptotic adiabatic regime  has been reached, $\Omega_{co} = \gamma \Omega_V$, and we we obtain
\be
\Omega_{co} = \frac{\gamma}{1 + \gamma} (1 - \Omega_c) \ .
\label{Omegaco}
\ee
Since $\Omega_c \sim 0.3$ today, and $\gamma \ll 1$ in the strong
coupling limit we are considering, we must have $\Omega_{co} \ll 1$
today.

The maximum and minimum lengthscales for the instability are still
given by Eqs.\ (\ref{Lmin2}) and (\ref{Lmax2}), but with $\rho$ replaced by
$\rho_{co}$.  Since $\rho_{\rm co}$ is approximately a constant in the
strong coupling limit, these lengthscales are also constants.
If the parameters of the model are chosen so that $\Omega_c \sim 1$
today, then
\be
{\cal L}_{\rm max} \sim H_0^{-1}, \ \ \ \ {\cal L}_{\rm min} \sim
\frac{H_0^{-1}}{\beta |C|} \ .
\label{scales0}
\ee
The evolution equations for the fractional density perturbations
$\delta_{\rm j} = \delta \rho_{\rm j} / \rho_{\rm j}$
in the adiabatic limit on subhorizon scales
are given by
\be
{\ddot \delta}_{\rm j} + 2 H {\dot \delta}_{\rm j} - 4\pi \sum_k
G_{{\rm j}{\rm k}}
\rho_{\rm k} e^{\alpha_{\rm k}} \delta_{\rm k}=0 \ ,
\ee
where the effective Newton's constants are given by Eq.\
(\ref{Gformula}) with $\alpha_b$ set to zero.  Writing this out
explicitly we obtain
\bea
{\ddot \delta}_c + 2 H {\dot \delta}_c &=& \frac{1}{2 \mpl^2} \rho_c \delta_c +
\frac{1}{2 \mpl^2} e^\alpha \rho_{co} \delta_{co}, \\
{\ddot \delta}_{co} + 2 H {\dot \delta}_{co} &=& \frac{1}{2 \mpl^2}
\rho_c \delta_c \nonumber \\
&&+
\frac{1}{2 \mpl^2} \left[ 1 + \frac{2 \beta^2 C^2}{1 + \frac{m_{\rm
        eff}^2 a^2}{k^2}} \right] e^\alpha \rho_{co}
\delta_{co} \ . \nonumber \\
\label{uns}
\eea
The condition for the instability to operate is that
the timescale associated with the second
term on the right hand side of Eq.\ (\ref{uns}) be short compared with
$H^{-1}$, or
\be
\frac{\beta^2 C^2 k^2}{m_{\rm eff}^2 a^2} \rho_{co} e^\alpha \gg H^2
\mpl^2 \ .
\label{condt4}
\ee
Now the effective mass for this model is given by
$m_{\rm eff}^2 = \beta^2 C^2 \mpl^{-2} (V + e^\alpha \rho_{co}) = 3
\beta^2 C^2 \mpl^{-2} H^2 (\Omega_V + \Omega_{co})$.  Substituting this
into Eq.\ (\ref{condt4}) and using Eq.\ (\ref{Omegaco}) gives the criterion
$k/(a H) \gg 1$.  Therefore
the instability should operate whenever modes are inside the horizon
and in the range of scales (\ref{scales0}).

\begin{figure}[t]
\begin{center}
\includegraphics[width=3.5in]{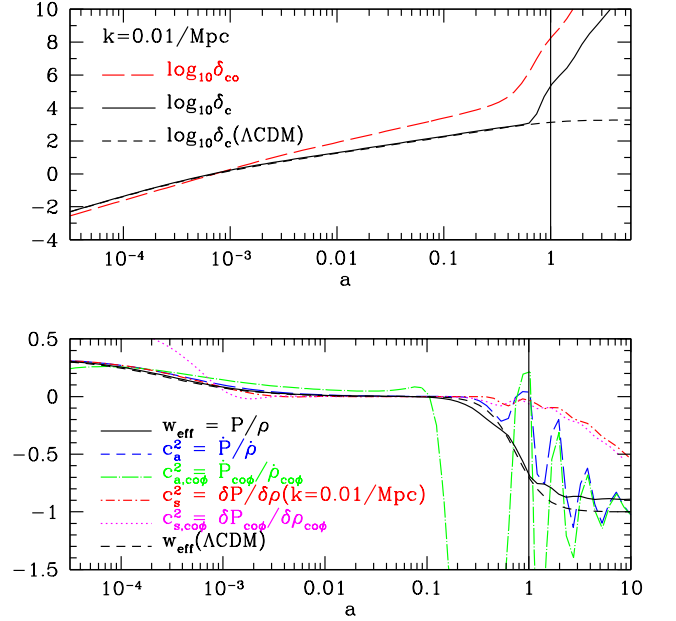}
\caption{[Bottom] The two component coupled dark energy (CDE)
  model, with exponential potential and coupling, with $\lambda =2$ and coupling $C=-20$ with $H_0=70 \, {\rm km} \, {\rm s}^{-1}\, {\rm Mpc}^{-1}$, $\Omega_{b}=0.05$,
  $\Omega_{c}=0.2$, $\Omega_{co}=0.05$, and $\Omega_{V}=0.70$. At late times the scalar field finds the adiabatic minimum with asymptotic equation of state, and sound speed $= -1/(1+\gamma) = -0.89$, able to reproduce a viable background evolution consistent   with supernovae, CMB angular diameter distance and BBN expansion
  history constraints. The figure shows the evolution of the effective equation of state, $w_{eff} = P_{tot}/\rho_{tot}=(2/3) (d\ln t / d\ln a) -1,$ (black full line), the adiabatic speed of sound, $c_{a}^{2}=\dot{P}/\dot{\rho}$ for all components (blue long dashed line) and for the coupled components only (green dot long dashed line), and effective speed of sound for $c_s^2=\delta P/\delta\rho$ at $k=0.01/Mpc$ for all components (red dot-dashed line)  and for the coupled components alone (magenta dotted line). The effective equation of state for a comparable $\Lambda$CDM model with $\Omega_{c}=0.25$, $\Omega_b=0.05$ and $\Omega_{\Lambda}=0.7$ is also shown (black dashed line). [Top] The growth of the fractional over-density $\delta=\delta\rho/\rho$ for $k=0.01/Mpc$ for the coupled CDM component, $\delta_{co}$, (red long dashed line) and uncoupled component, $\delta_{c}$, (black full line) in comparison to the growth for the $\Lambda$CDM model (black dashed line). At late times the adiabatic behavior triggers a dramatic increase in the rate of growth of both uncoupled and coupled components, leading to structure predictions inconsistent with observations.\label{fig1}}
\end{center}
\end{figure}

These expectations are confirmed by numerical integrations.
In figure \ref{fig1}  we present a numerical analysis
of such a two component model. We consider an exponential potential with $\lambda =2$ and strong
coupling with $C=-20$, and typical cosmological parameters are assumed,
$H_0=70 \, {\rm km} \, {\rm s}^{-1}\,{\rm Mpc}^{-1}$,
$\Omega_{b}=0.05$, $\Omega_{c}=0.2$, $\Omega_{co}=0.05$, and
$\Omega_{V}=0.7$. We fix initial conditions of $\phi/\mpl=10^{-3}$ and
$\dot\phi=0$ at $a=10^{-10}$, although the dynamical attractor renders
the final evolution largely insensitive to these choices, we have checked that $\phi/\mpl(a=10^{-10})=10^{-30}-1$ give the same evolution. We neglect the effect of the coupled CDM component peculiar velocity in the initial conditions (it is many orders of magnitude smaller than  the density perturbation) and assume
that the coupled  and uncoupled CDM components have the same initial
fractional density perturbations $\delta_c=\delta_{co}$, fixed by the
usual adiabatic initial conditions. As shown in figure \ref{fig1}, the
background evolution is entirely consistent with a $\Lambda$CDM like
scenario, with $w_{\rm eff}=-0.69$ today
and an asymptotic equation of state at
late times, given by (\ref{weff1}), $w_{\rm eff} = -1/(1+\gamma)=-0.89$. The large
coupling drives the evolution to an adiabatic regime at late times,
with an adiabatic sound speed $c_{a}^{2}\rightarrow -1/(1+\gamma)$ as
in (\ref{cs21}). This drives a rapid growth in over-densities once in the adiabatic regime so that although consistent with structure observations at early times, they are inconsistent once the accelerative regime has begun.

In summary, these models provide a class of theories for which the
background cosmology is compatible with observations, but which are
ruled out by the adiabatic instability of the perturbations.

\subsection{Chameleon models}

Next we study the so-called chameleon models \cite{Khoury:2003aq,Khoury:2003rn}
defined by the potential
\be
V(\phi) = \lambda M^4 \left( \frac{M }{\phi} \right)^n \ ,
\ee
where $M$ is a mass scale and $n>0$ and $\lambda$ are dimensionless constants,
together with the coupling function~(\ref{constantcoupling}).
In these models it has been previously shown that the adiabatic regime
is achieved in static solutions describing macroscopic bodies like
the Earth, and that cosmological solutions in the adiabatic regime
provide good models of dark energy \cite{Brax:2004qh,Brax:2004px,Brax:2005ew}.
We now study under what conditions these models are
unstable.

The effective potential is, from Eq.\ (\ref{phieqn}),
\be
V_{\rm eff}(\phi,\rho) = \lambda M^4 \left( \frac{M }{\phi} \right)^n
 + e^{-\beta C \phi/\mpl} \rho \ ,
\ee
and solving for the local minimum of this potential yields
the relation between $\phi$ and $\rho$ in the adiabatic regime:
\be
x^{n+1} e^x = \frac{\rho_{\rm crit}}{\rho} \ ,
\label{xeqn}
\ee
Here $x = -\beta C \phi_c(\rho)/\mpl$ is dimensionless and the critical density is
\be
\rho_{\rm crit} = n \lambda M^4 \left(\frac{-\beta C M}{\mpl}\right)^n \ .
\ee
As before the existence of a local minimum in the effective potential requires
$C$ to be negative.
We shall restrict attention to the regime
\be
\rho \gg \rho_{\rm crit}
\label{highdensity}
\ee
since for models of dark energy $\rho_{\rm crit}$ will be of order the
present day cosmological density.  In this
regime the solution to Eq.\ (\ref{xeqn}) is approximately
\be
x \approx \left( \frac{\rho_{\rm crit} }{\rho} \right)^{\frac{1}{n+1}}.
\ee
The corresponding effective mass parameter is
\be
m_{\rm eff}^2 = (x + n + 1) n \lambda M^2 \left( \frac{-\beta C M}{x \mpl} \right)^{n+2},
\label{mass2}
\ee
which in the regime (\ref{highdensity}) simplifies to
\be
m_{\rm eff}^2 = (n+1) (-\beta C)^2 \mpl^{-2} \rho_{\rm crit} \left( \frac{\rho}{\rho_{\rm crit}}
\right)^{\frac{n+2}{n+1}}.
\ee

Next, we compute the sound speed squared.  Using Eq.\
(\ref{soundspeed}) we obtain
\be
\frac{1}{c_s^2}  = -1 -  \frac{n+1}{(-\beta C)} \frac{\mpl}{\phi},
\label{cs213}
\ee
and since $C$ is negative, we see that this
model is always unstable in the adiabatic regime.
Inserting the effective mass (\ref{mass2}) and the sound speed squared (\ref{cs213})
into Eq.\ (\ref{range}) we obtain the range of spatial scales ${\cal
  L}$ over which
the instability operates for a given density $\rho$:
\be
{\cal L}_{\rm min}(\rho)^2 \ll {\cal L}^2 \ll
(\beta C)^2  {\cal L}_{\rm min}(\rho)^2,
\label{range2}
\ee
where
\be
{\cal L}_{\rm min}^2(\rho) = \frac{\mpl^2}{ (n+1) (\beta C)^2 \rho_{\rm crit}} \left(
  \frac{ \rho_{\rm crit} }{\rho} \right)^{\frac{n+2}{n+1}}.
\ee
We see that the range of unstable lengthscales is non-empty only if
\be
\beta |C| \gg 1,
\label{ad34}
\ee
which as before is equivalent to the strong coupling condition (\ref{strongcoupling}).
Note that the first of the two inequalities in
Eq.\ (\ref{range2}) is equivalent to the ``thin-shell condition'' of Ref.\
\cite{Khoury:2003rn}
when the background value of the scalar field can be neglected.

As for the exponential models of Sec.\ \ref{sec:exponential}, the
effective equation of state $w_{\rm eff}$ is close to $-1$
in the adiabatic regime for large coupling, assuming $\rho \gg \rho_{\rm crit}$.
Therefore these models do not give an acceptable background cosmology for the matter dominated era
in their adiabatic regime.  However, one can construct two component
models analogous to those in Sec.\ \ref{sec:twocomponent} using the
chameleon potential.  Those models give an acceptable background
cosmology, but are then ruled out by the adiabatic instability.

\subsection{Mass Varying Neutrino (``MaVaN'') models }
\label{sec:MaVaN}

The impact of adiabatic instabilities has been discussed extensively in the context of MaVaN models, in which the light mass of the neutrino and the recent accelerative era are twinned together through a scalar field coupling \cite{Fardon:2003eh,Kaplan:2004dq,Fardon:2005wc}.  The adiabatic instability was shown to be a concern in these models with the implication of forming compact localized regions of neutrinos after undergoing dramatic adiabatic collapse \cite{Afshordi:2005ym}.

The action for the MaVaN models is of the form (\ref{eq:action10}) with $p=0$.
Therefore the analysis in the earlier sections of this paper, which
assumed $p=2$, does not directly apply to these models.  However it is
straightforward to generalize our analysis to cover this case.
Consistent with the general discussion in sections
\ref{adiabatic}-\ref{instability}, the instability is present in these
theories if the coupling is strong. Here we discuss examples of
models which evade or are subject to the instability on cosmological
scales.

Recently,  Ref.\ \cite{Bjaelde:2007ki} discussed a MaVaN scenario with a logarithmic potential and scalar dependent mass,
\bea
V(\phi) &= &V_0\log(1+\xi\phi) \
\\
\label{mavan1}
m_\nu(\phi) &=& m_{\nu 0}\left(\frac{\phi_*}{\phi}\right) \ ,
\eea
where $\xi$ is a constant and $m_{\nu 0}$ and $\phi_*$ are the current values of the neutrino mass and the scalar field respectively.
It was found in \cite{Bjaelde:2007ki} that this model can exhibit an instability in growth for an otherwise cosmologically viable background solution. We find, however, that this model can also allow stable solutions for identical fractional densities today as those studied in \cite{Bjaelde:2007ki}, for a wide range of parameter values. If the coupling, $m_\nu'/m_\nu$, is not large compared to $\mpl^{-1}$ at late times, the evolution never enters the adiabatic regime on cosmic scales. This translates in this model to $\phi_*$ not being significantly less than $\mpl$.

The MaVaN model evolves according to a coupled Klein Gordon equation,
\bea
\ddot{\phi} + 2H \dot\phi +a^2V'(\phi) &=& -a^2 (\rho'_\nu(\phi)-3P'_\nu(\phi)) \ .\ \ \ \ \ \ \label{KGnu}
\eea
Assuming no chemical potential, and $g$ spin states per neutrino species with momentum $p$ and mass $m$ ,the neutrino density and pressure are given by
\bea
a^4 \rho_\nu& = & {g (k_{B}T_{\nu}^{0})^{4} \over 2\pi^{2}} \int_{0}^{\infty} dq q^{2} (q^{2}+a^{2}\bar{m}_\nu^{2})^{1\over 2} f(q) \\
a^4 P_\nu & = & {g (k_{B}T_{\nu}^{0})^{4} \over 2\pi^{2}} \int_{0}^{\infty} dq q^{2} {q^{2}\over 3(q^{2}+a^{2}\bar{m}_\nu^{2})^{1\over 2}} f(q) \ \  \ \ \
\\
f(q)&\approx& \left[\exp\left({q}\right)+1\right]^{-1} \ ,
\eea
 with $q \equiv ap/ k_{B}T_{\nu}^{0}$ and $\bar{m}_\nu\equiv m_\nu c^{2}/k_{B}T_{\nu}^{0}$.

In the relativistic regime, with $\bar{m}\ll 1$, the potential is negligible and the driving term, on the right hand side of  (\ref{KGnu}), can be calculated by doing a Taylor expansion to first order in $\bar{m}_\nu$,
\bea
a^4 (\rho_\nu-3P_\nu)
& \approx &{g (k_{B}T_{\nu}^{0})^{4} \over 2\pi^{2}} \int_{0}^{\infty} dq q f(q) a^{2}\bar{m}_\nu^{2} \ \
\\
a^{2}(\rho_\nu-3P_\nu)
&\approx&  \frac{10}{7\pi^2}\bar{m}_\nu(\phi)^{2}\rho_0 \ ,
\eea
so that
\bea
a^{2}(\rho'_\nu-3P'_\nu) &\approx& -\frac{2}{\phi}(\rho_\nu-3P_\nu) a^{2} \ ,
\label{rhopres}
\eea
where $\rho_0\equiv 7\pi^2 g(k_BT_\nu^0)^4 /240$ would be the relativistic neutrino energy density per neutrino species today with temperature $T_{\nu}^{0}$.

Putting (\ref{rhopres}) into (\ref{KGnu}) and neglecting the potential
we find a power law attractor $\phi\propto \tau^{x}$ with $x=0.5$.
The normalization of $\phi$ is wholly specified in the attractor by (\ref{KGnu}). Writing $\phi = \phi_i(\tau/\tau_i)^{0.5}$, and $a\propto\tau^{p}$, with $p=2/(1+3w_{\rm eff})$, we find
\bea
\phi &=& \left[\frac{80\rho_0m_{\nu 0}^{2}\phi_*^{2}}{7\pi^2(4p-1)}\right]^{0.25}\tau^{0.5} \ .
\label{phinorm}
\eea

When the neutrino is non-relativistic, if we again neglect the potential,
\bea
(\rho_\nu-3P_\nu) &=&  \frac{3H_{0}^{2}\mpl^2 \Omega_\nu}{a^3} \left(\frac{\phi_*}{\phi}\right)
\\
a^2(\rho'_\nu-3P'_\nu) &=&-\frac{3H_{0}^{2}\mpl^2 \Omega_\nu\phi_*}{a} \left(\frac{1}{\phi^{2}}\right) \ .
\eea
The Klein-Gordon equation has a solution $\phi \propto \tau^x$ with $x=(2-p)/3$, tending towards a cessation of growth in $\phi$ in the matter dominated era.
\begin{figure}[t]
\begin{center}
\includegraphics[width=3.6in]{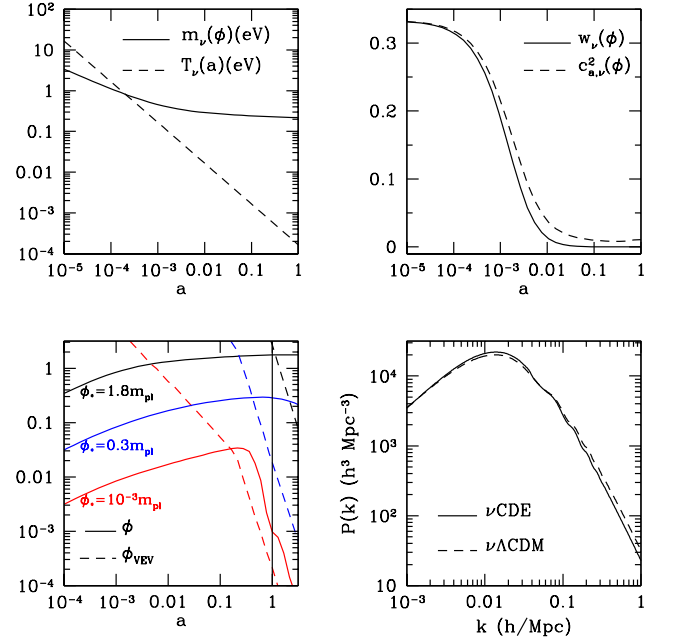}
\caption{[Top panels] Evolution of $m_\nu(\phi)$ and neutrino temperature (left), and associated equation of state and adiabatic sound speed (right) for MaVaN model described in the text with $m_{\nu 0}=0.312eV$ and $\phi_*=1.8\mpl$. [Bottom left panel] Scalar field evolution in the MaVaN scenario, for 3 values of $\phi_*\sim 10^{-3}\mpl$, $0.3\mpl$ and $1.8\mpl$ (full lines) showing the $\tau^{0.5}$ attractor while the neutrino is relativistic allowing late time evolution to be independent of initial conditions. The vacuum expectation value of the scalar field, if the field becomes adiabatic, (dashed lines) is also shown. For $\phi_* \gtrsim 10^{-2}\mpl$ the scalar field does not enter an adiabatic era on cosmological scales before now, and growth of perturbations remains well-behaved. For smaller $\phi_*$, for example $\phi_*\sim 10^{-3}\mpl$ shown, the evolution is adiabatic at late times, similar to that discussed in \cite{Bjaelde:2007ki}. [Bottom right panel] The resulting matter power spectrum from the coupled dark energy (CDE) model with $\phi_* = 1.8\mpl$ and $\Omega_\nu=0.02$ (full line) is very similar to that for $\Lambda$CDM (dashed line) with the same baryon fraction and $H_0$, $\Omega_m=0.3$, $\Omega_\nu=0.02$ when normalized at large scales.}\label{fig2}
\end{center}

\end{figure}

If the neutrinos are non-relativistic (but, of course, assuming that
they decoupled when they were relativistic)
\bea
(\rho_\nu-3P_\nu) &\approx& N_\nu \frac{n_0}{a^3} m_{\nu 0} \left(\frac{\phi_*}{\phi}\right)
\\
& \approx& \frac{180\zeta(3)}{7\pi^4}\frac{\rho_0}{a^{3}} N_\nu \bar{m}_{\nu0}\left(\frac{\phi_*}{\phi}\right) \ .
\eea
At late times, the complete effective potential is relevant, for which there exists a minimum at positive $\phi$ given by
\bea
\frac{\phi_{VEV}}{\phi_*}  = \frac{1}{a^{3}}\frac{\Omega_{\nu}}{\Omega_{pot}}\log(1+\xi\phi_*) \ .
\eea

We modified CAMB \cite{Lewis:2002ah} to investigate the evolution numerically.
In figure \ref{fig2} we show the scalar field and neutrino mass evolution for $\xi = 10^{20}\mpl^{-1}$, $\phi_*=1.8\mpl$, and $m_{\nu 0}=0.312eV$ (giving $\Omega_\nu=0.02$), $\Omega_b=0.05$, $\Omega_c=0.23$, $H_0=70 $kms$^{-1}$Mpc$^{-1}$.  In the bottom left hand figure, the numerical evolution of scalar field is shown for $\phi_*=1.8\mpl$ along with two smaller values $\phi_*\sim 10^{-3}\mpl$ and $0.3\mpl$ for which the minimum of the effective potential is steeper. The background evolution obeys the $\phi\sim\tau^{0.5}$ attractor and normalization in (\ref{phinorm}), rendering it largely independent of the initial conditions. When the neutrino becomes non-relativistic the evolution slows and for small values of $\phi_*$ starts to track the VEV, as discussed in \cite{Bjaelde:2007ki}. However for larger values of $\phi_*$, the VEV is not reached until later times, and in addition, the effective potential minimum is shallow enough that the scalar field does not get fixed at the VEV automatically, enabling well-behaved growth. The bottom right hand figure shows the resultant matter power spectrum, which is very similar to a fiducial $\Lambda$CDM model with the same Hubble factor, baryon and neutrino density today when normalized to the same amplitude at large scales.

\section{Conclusions}
\label{conclusions}

If dark energy and dark matter are to fit into a coherent fundamental
physics framework then there are likely couplings between them.
Such couplings may have far reaching macroscopic implications on scales
ranging from the solar system up to cosmological horizon. As we have
discussed, these effects may lead to observationally distinctive
characteristics, which may allow us to tease out the nature of the
dark sector. However they may also give rise to catastrophic
instabilities with which we may constrain the class of physically
viable dark energy models.

In this paper we have  considered such theories in which there exists a nontrivial coupling between the
dark matter sector and the sector responsible for the acceleration of the universe.

We have comprehensively analyzed an instability -- characterized by a
negative sound speed squared of
an effective coupled dark matter/dark energy fluid -- that exists
whenever such theories enter an adiabatic regime in which the scalar
field faithfully tracks the minimum of the effective potential, and
the coupling strength is strong compared to gravitational strength.
The adiabatic regime occurs when the relaxation time scale associated with
the scalar field is much shorter than the Hubble time. We have
demonstrated how this instability can be viewed from the
alternative perspectives of the kinetic theory of dark matter and as a
Jeans instability associated with modified Newton's constants.

We have established the conditions under which the adiabatic instability occurs, finding  a condition on the coupling, $|\alpha'|\mpl\gg1$, and have identified the time and length scales over which the instability is active in a given setting governed by the matter density.  These length scales can differ greatly dependent on whether one is considering galactic or cosmic densities.

Our work builds on previous analyses  of MaVaN
\cite{Fardon:2003eh,Kaplan:2004dq,Fardon:2005wc,Afshordi:2005ym,Bjaelde:2007ki},
chameleon \cite{Khoury:2003aq,Khoury:2003rn}, and general coupled
scenarios \cite{Kaplinghat:2006jk}. Our numerical analyses of coupled
CDM and MaVaN models reinforce our analytic findings. We show that the
regime of adiabatic behavior agrees with the predictions based on coupling strength and length scale
conditions mentioned above.  In the appropriate limits, our results
reduce to the previous findings in a number of cases, while in several
other cases we have provided corrected results. In particular,  we
have shown that stable MaVaN models exist that evade the instability,
if $m_\nu'/m_\nu$ is not large compared to $\mpl$.

\acknowledgments

We would like to thank Ole Bjaelde,  Anthony Brookfield, Steen
Hannestad, Carsten Van der Bruck, Ira Wasserman and Christoph
Wetterich for useful discussions in the course of this work.
We thank an anonymous referee for some helpful comments.
RB's work is supported by National Science Foundation (NSF) grants AST-0607018 and PHY-0555216, EF's by
NSF grants PHY-0457200 and PHY-0555216, and MT's by NSF grant
PHY-0354990 and by the Research Corporation.

\appendix

\section{Effect of normal matter on instability}
\label{visible}

In the body of this paper
we have neglected the effect of baryons and the other matter
species, assuming that their
density is much smaller than that of the dark matter.
A natural
question is does the instability persist in regimes where the density
of visible matter
density is comparable to or larger than that of the dark matter -- one
might expect the instability to be killed by
the pressure of the visible matter.  In general the answer
depends on the coupling function $\alpha_b$ in the action
(\ref{action0}), which governs the coupling of the scalar field $\phi$
to normal matter.  The instability persists in two specific cases, (i)
$\alpha_b =0$ where normal matter is uncoupled from $\phi$, and
(ii) $\alpha_b = \alpha_c$, the case usually considered in which
dark matter and normal matter couple the same way to the scalar field,
respecting the equivalence principle.

To derive this result, we generalize the derivation given in the body of the paper
from a single, pressureless fluid to a set of $N$ non-interacting
fluids, each of which can have a pressure.
We start from the action (\ref{action0}),
and we define the Jordan frame metric that the ${\rm
  j}$th sector couples minimally to by
\be
{\bar g}_{{\rm j}\,ab} = e^{2 \alpha_{\rm j}(\phi)} g_{ab} \ .
\label{Jordan1}
\ee
The stress energy tensor ${\bar T}_{{\rm j}}^{\,ab}$ of the ${\rm
j}$th sector is defined by
\be
S_{\rm j}[ {\bar g}_{{\rm j}\,ab} + \delta {\bar g}_{{\rm j}\,ab}, \Psi_{\rm
    j}] = S_{\rm j}[ {\bar g}_{{\rm j}\,ab}, \Psi_{\rm j}] + \frac{1}{2}
\int d^4 x \sqrt{- {\bar g}_{\rm j}} {\bar T}_{{\rm j}}^{\,ab}
\delta {\bar g}_{{\rm j}\,ab} \ ,
\ee
and we assume it has the form of a perfect fluid
\be
{\bar T}_{{\rm j}}^{\,ab} = ( {\bar \rho}_{\rm j} + {\bar p}_{\rm
  j}) {\bar u}_{\rm j}^{\,a} {\bar u}_{\rm j}^{\,b} + {\bar
  p}_{\rm j} {\bar g}_{\rm j}^{\,ab} \ .
\ee
Here ${\bar u}_{\rm j}^{\,a}$ is the 4-velocity which is normalized
according to ${\bar g}_{{\rm j}\,ab} {\bar u}_{\rm j}^{\,a}
{\bar u}_{\rm j}^{\,b} = -1$.
This perfect fluid assumption requires that the matter in the jth
sector be barotropic, ie that its pressure be determined uniquely by its density,
which in many regimes is a good approximation.

The equation of motion of the ${\rm j}$th fluid is
\be
{\bar \nabla}_{{\rm j}\,a} {\bar T}_{\rm j}^{\,ab} = 0 \ ,
\ee
where ${\bar \nabla}_{{\rm j}\,a}$ is the derivative operator
determined by the metric ${\bar g}_{{\rm j}\,ab}$.
The components of this equation perpendicular to and parallel to the
4-velocity are
\be
{\bar a}_{\rm j}^a = - \frac{1}{{\bar \rho}_{\rm j} + {\bar p}_{\rm
    j}} ({\bar g}_{\rm j}^{ab} + {\bar u}_{\rm j}^a {\bar u}_{\rm
  j}^b)  {\bar \nabla}_{{\rm j}\,b} {\bar p}_{\rm j} \ ,
\ee
where ${\bar a}_{\rm j}^a = {\bar u}_{\rm j}^{\,b} {\bar \nabla}_{{\rm
    j}\,b} {\bar u}_{\rm j}^{\,a}$ is the Jordan frame 4-acceleration,
and
\be
{\bar \nabla}_{{\rm j}\,a} \left[ ({\bar \rho}_{\rm j} + {\bar p}_{\rm
    j}) {\bar u}_{\rm j}^{\,a} \right] = {\bar u}_{\rm j}^{\,a} {\bar
  \nabla}_{{\rm j}\,a} {\bar p}_{\rm j} \ .
\ee
Next, we conformally transform these fluid
equations
using Eq.\ (\ref{Jordan1}) to
write them in terms
of the Einstein frame metric $g_{ab}$ and the Einstein-frame
normalized 4-velocities $u_{\rm j}^{\,a}=e^{\alpha_{\rm j}}{\bar u}_{\rm j}^{\,a}$ which satisfy $g_{ab} u_{\rm
  j}^{\,a} u_{\rm j}^{\,b} =-1$.  This gives
\be
a_{\rm j}^{\,a} = -
 (g^{ab} + u_{\rm j}^a u_{\rm j}^b)  \left[ \frac{ \nabla_b {\bar p}_{\rm j}}
  {{\bar \rho}_{\rm j} + {\bar p}_{\rm j}} + \nabla_b \alpha_{\rm j}\right] \ ,
\label{fluid1}
\ee
where $a_{\rm j}^{\,a} = u_{\rm j}^{\,b} \nabla_b u_{\rm j}^{\,a}$ and
\be
e^{-3 \alpha_{\rm j}} \nabla_a \left[ e^{3 \alpha_{\rm j}} ({\bar
      \rho}_{\rm j} + {\bar p}_{\rm j} ) u_{\rm j}^{\,a} \right] =
  u_{\rm j}^a \nabla_a {\bar p}_{\rm j} \ .
\label{fluid2}
\ee

Next, we consider the non-relativistic limit of the fluid equations
(\ref{fluid1}) and (\ref{fluid2})
together with the adiabatic limit of
the scalar field equation (\ref{eq:scalar10}).  We also
neglect self-gravity in the Einstein frame, taking $g_{ab} \approx \eta_{ab}$.  This
approximation should be valid on sufficiently small spatial scales.
We get
\be
\frac{\partial}{\partial t} \left( e^{3 \alpha_{\rm j}} {\bar
  \rho}_{\rm j} \right) + {\bf \nabla} \cdot \left( e^{3 \alpha_{\rm
    j}} {\bar \rho}_{\rm j} {\bf v}_{\rm j} \right) =0 \ ,
\ee
\be
\frac{ \partial {\bf v}_{\rm j}} {\partial t} + ( {\bf v}_{\rm j}
\cdot {\bf \nabla} ) {\bf v}_{\rm j} = - \frac{ {\bf \nabla} {\bar
    p}_{\rm j}}{{\bar \rho}_{\rm j}} - {\bf \nabla} \alpha_{\rm j} \ ,
\label{euler1}
\ee
where $\alpha_{\rm j} = \alpha_{\rm j}(\phi)$ and $\phi = \phi({\bar
  \rho}_{\rm k})$ is given by the equation
\be
V'(\phi) = - \sum_{\rm j} \alpha'_{\rm j}(\phi) e^{4 \alpha_{\rm j}(\phi)}
{\bar \rho}_{\rm j} \ .
\label{scalar12}
\ee

Next, we switch to using the rescaled density variables
\be
\rho_{\rm j} = e^{3 \alpha_{\rm j}} {\bar \rho}_{\rm j} \ .
\ee
We linearize the resulting equations about the background solution
$\rho_{\rm j} = \rho_{{\rm j}0} = $ constant, ${\bf
  v}_{\rm j} = 0$, $\phi = \phi_0$.  In other words we assume that the
dark matter and
visible matter densities are constant and that the fluids are not
in relative motion.  We define $\alpha_{{\rm j}0} = \alpha_{\rm
  j}(\phi_0)$ and $\alpha'_{{\rm j}0} = \alpha'_{\rm j}(\phi_0)$,
and look for a solution
\begin{eqnarray}
\rho_{\rm j} &=& \rho_{{\rm j}0} + \delta \rho_{\rm
  j},\\
\phi &=& \phi_0 + \delta \phi \ .
\end{eqnarray}
From Eq.\ (\ref{scalar12}) we obtain that $\delta \phi = \sum_j \chi_{\rm j}
\delta \rho_{\rm j}$, where
\be
\chi_{\rm j} = - \frac{ \alpha'_{{\rm j}0} e^{\alpha_{{\rm j}0}} }
{V''(\phi_0) + \sum_{\rm k} \left[ \alpha''_{{\rm k}0} +
    (\alpha'_{{\rm k}0})^2\right] e^{ \alpha_{{\rm k}0}}
    \rho_{{\rm k}0}} \ .
\label{chidef}
\ee
Next we assume that all the variables are proportional to $\exp
\left[i {\bf k} \cdot {\bf x} - i \omega t \right]$, and that the
velocities are proportional to ${\bf k}$.  This assumption will
yield one particular set
of modes of oscillation of the coupled fluids, but these modes
will contain the instability if it is present.  The resulting
eigenvalue equation for $\omega^2$ is
\be
\omega^2  \delta \rho_{\rm j} = {\bf k}^2 \Gamma_{{\rm j}{\rm k}} \delta
        \rho_{\rm k} \ ,
\label{evaleqn}
\ee
where the matrix $\Gamma_{{\rm j}{\rm k}}$, which plays the role of
the effective squared sound speed, is
\be
\Gamma_{{\rm j}{\rm k}} = {\bar c}_{\rm j}^2 \delta_{{\rm j}{\rm k}} +
(1 - 3 {\bar c}_{\rm j}^2) \rho_{{\rm j}0} \alpha'_{{\rm j}0}
\chi_{\rm k} \ .
\label{Gamma}
\ee
Here
\be
{\bar c}_{\rm j}^2 = \frac{d {\bar p}_{\rm j}}{d {\bar \rho}_{\rm j}}
\ee
is the physical (Jordan frame) squared sound speed of the ${\rm j}$th fluid.

From the eigenvalue equation (\ref{evaleqn}), it follows that
the system will be stable if and only if
all of the eigenvalues of
the matrix $\Gamma_{{\rm j}{\rm k}}$ are real and positive.
For the case of a single fluid with ${\bar c}_{\rm j}^2=0$, this criterion
reduces to the criterion (\ref{unstablec}) derived in the body of the paper.
In general, there is a competition between the positive squared
sound speeds of the fluids in the first term in Eq.\ (\ref{Gamma}), and the
(possibly) negative squared sound speeds coming from the interaction
with the scalar field given in the second term.

We now specialize to two fluids, a dark matter fluid with zero sound
speed, and a fluid describing the visible matter.  Taking ${\rm j}
= 1 = c$ for the dark matter (CDM) and ${\rm j} = 2 = b$ for
the visible matter (baryons), and defining
\be
\nu_{\rm j}  = (1 - 3 {\bar c}_{\rm j}^2) \rho_{{\rm j}0} \alpha'_{{\rm j}0}
\label{nudef}
\ee
gives for the matrix
\begin{equation}
\Gamma_{{\rm j}{\rm k}} = \left( \begin{array}{cc}
\nu_{c} \chi_c &
\nu_{c} \chi_b   \\
\nu_{b} \chi_c   &
{\bar c}_b^2 + \nu_b \chi_b
\end{array} \right) \ .
\label{matrix1}
\end{equation}
Now if the system is stable, then both of the eigenvalues $\lambda_1$
and $\lambda_2$ of the
matrix must be real and nonnegative, and hence both the determinant
$\lambda_1 \lambda_2$ and
the trace $\lambda_1 + \lambda_2$ of the matrix must be nonnegative.
Conversely, if either the determinant or the trace of the matrix is
negative, then the system is unstable.
The determinant is
\begin{eqnarray}
{\rm det} \, {\bf \Gamma} &=& {\bar c}_b^2 \nu_c
 \chi_c
= - \frac{ ( \alpha'_c)^2 e^{\alpha_c} {\bar c}_b^2 \rho_c}
{\Upsilon
} \ , \ \ \ \
\label{detGamma}
\end{eqnarray}
where
\bea
\Upsilon &=& V'' + \left[ \alpha''_c + (\alpha'_c)^2 \right]
e^{\alpha_{c}} \rho_c
\nonumber \\ &&
+ \left[ \alpha''_b + (\alpha'_b)^2 \right] e^{\alpha_b} \rho_b
\label{Upsilondef}
\eea
and where we have used Eqs.\ (\ref{chidef}) and (\ref{nudef}).
In Eqs.\ (\ref{detGamma}) and (\ref{Upsilondef}), $\phi$ can be taken
to be the function of $\rho_{c}$ and $\rho_b$ given by [cf.\ Eq.\
(\ref{fluid2})]
\be
V'(\phi) = - \alpha'_c(\phi) e^{\alpha_c(\phi)}
\rho_c - \alpha'_b(\phi) e^{\alpha_b(\phi)}
\rho_b \ .
\label{Phieqn4}
\ee
Equations (\ref{detGamma}) -- (\ref{Phieqn4}) allow us to determine
which values of $\rho_c$ and $\rho_b$ satisfy the sufficient condition
${\rm det} \, {\bf \Gamma} < 0$ for instability,
given the functions $V(\phi)$, $\alpha_c(\phi)$ and
$\alpha_b(\phi)$.  Note that $\alpha_c(\phi)$ was denoted by
$\alpha(\phi)$ in the body of the paper.

We now consider two special cases.  If $\alpha'_b=0$, so that
the visible matter is not coupled to the scalar field, then
eigenvalues of the matrix ${\bf \Gamma}$ are just ${\bar c}_{b}^2$ and
$\nu_c \chi_c$.  The second of these
eigenvalues coincides with the effective squared sound speed computed above in
Eq.\ (\ref{soundspeed}).  Thus we recover the results of the body of
the paper.

The second special case is when $\alpha_b = \alpha_c$.
In this case the instability criterion ${\rm det} \, {\bf \Gamma}<0$ again reduces to the criterion
(\ref{soundspeed}) computed earlier for a single fluid, with the
modification that the effective sound speed squared is now a function
of the total density $\rho = \rho_c + \rho_b$
rather than just of $\rho_c$.  If this
total density is in the unstable regime, then the instability will
persist despite the pressure of the visible matter.

Finally, one can also consider the effect of the scalar field on just
the baryons, neglecting the dark matter.  The sound speed of the
baryons gets an additional term, but it is a small correction unless
the coupling is large, $\alpha_b^\prime \mpl \gg 1$, and observational
tests of general relativity in the Solar System
require $\alpha_b^\prime \mpl \ll 1$.

\section{Kinetic theory treatment of instability}
\label{kinetic}

In this appendix we describe dark matter using the collisionless
Boltzmann equation, and specialize to the non-relativistic regime and
to lengthscales small enough that self-gravity can be neglected.
We show, first, that the instability is generic, occurring for any
velocity distribution function with sufficiently small velocity
dispersion, and, second, that for Maxwellian distributions
the fluid is stabilized by free streaming once the velocity dispersion
becomes larger than a critical value.

Starting from the general
action (\ref{action0}), we specialize to the non-relativistic limit
neglecting self-gravity ($g_{ab} \approx \eta_{ab}$), which will
be valid on sufficiently small spatial scales.  The Jordan-frame
metric is then
$$
ds^2 = e^{2 \alpha} (-dt^2 + d{\bf x}^2) \ .
$$
We assume that dark matter is composed of non-interacting,
non-relativistic particles of mass $\mu$.
We denote by $f(t,{\bf x},{\bf v})$ the one-particle distribution function,
normalized so that the number of particles in the volume element
$d^3x$ and in the velocity region $d^3v$ is
$$
f(t,{\bf x},{\bf v})  d^3x d^3v \ .
$$
Now the physical (Jordan-frame)
volume element is $e^{3 \alpha} d^3 x$,
so the Jordan frame mass density ${\bar \rho}$ [cf.\
Eq.\ (\ref{ee0}) above] is given by
$
{\bar \rho} = \mu e^{-3 \alpha} \int d^3v f.
$
From Eq.\ (\ref{newdensity}) the rescaled density variable $\rho$ is then given
by
\be
\rho = \mu \int d^3v f \ .
\label{kineticrho}
\ee
With this notation the collisionless Boltzmann equation is [cf.\ Eq.\
(\ref{euler1}) above with ${\bar p}_{\rm j}=0$]
\be
\frac{\partial f}{\partial t} + v^i \frac{\partial f}{\partial x^i} =
\frac{\partial \alpha}{\partial x^i} \frac{\partial f}{\partial v^i} \ .
\ee
In the adiabatic limit, the variable $\alpha$ in this equation is
given by $\alpha = \alpha(\phi)$, where $\phi$ is given in terms of
$\rho$ by Eq.\ (\ref{eq:alg}), and $\rho$ is given in turn in terms
of $f$ by Eq.\ (\ref{kineticrho}).

We now linearize the Boltzmann equation about a homogeneous background
solution,
taking
$
f = f_0({\bf v}) + \delta f(t,{\bf x},{\bf v}).
$
and taking $\delta f \propto \exp[- i \omega t + i {\bf k} \cdot {\bf
  x}]$.  Using the identity $d\alpha/d\rho = c_s^2/\rho$ from Eq.\
(\ref{soundspeed}), and integrating over ${\bf v}$, gives the dispersion
relation
\be
\int d^3 v \frac{ {\bf k} \cdot \frac{\partial f_0}{\partial {\bf
      v}}}{\omega - {\bf k} \cdot {\bf v}} = - \frac{\rho_0}{\mu c_s^2} \ .
\ee
Here $\rho_0$ is the background density.
Without loss of generality we take ${\bf k}$ to be in the
$z$-direction, assuming $f_0$ is isotropic, and we define
\be
F(v) = \frac{\mu}{\rho_0} \int_{-\infty}^\infty dv_x
\int_{-\infty}^\infty dv_y \, f_0(v_x,v_y,v) \ .
\ee
This distribution function is normalized so that $\int dv F =1$, and
the dispersion relation can now be written as
\be
1 + c_s^2 \int dv \, \frac{F'(v)}{\omega/k - v} =0 \ .
\label{dispersion}
\ee
Given the distribution function $F(v)$ we can solve this equation to obtain
$\omega = \omega(k)$, which is complex in general.

The integrand in Eq.\ (\ref{dispersion}) contains a singularity at $v =
\omega/k$.  The derivation given here is incomplete since it does not
provide a specification for how to deal with the singularity.
However, just as in plasma physics \cite{Landau}, it is possible to give an
alternative derivation based on Laplace transforms.  That alternative derivation yields the
following specification: the integral over $v$ must be taken over the
Landau contour, which runs along the real axis if ${\rm Im}(\omega) > 0$, but
dips below the real axis to encircle the pole at $v = \omega/k$ if
${\rm Im}(\omega) \le 0$.

We now specialize to the adiabatic regime where $c_s^2 < 0$, and we
write $c_s^2 = - \beta^2$.  We write the complex frequency in terms of
its real and imaginary parts, $\omega = \omega_r +  i \omega_i$, and
we look for a solution $\omega = \omega(k)$ of the dispersion relation
(\ref{dispersion}) with $\omega_i > 0$, corresponding to an unstable
mode.  For such a solution the Landau contour is along the real
axis which simplifies the analysis.

If the velocity dispersion of the distribution $F(v)$ is small, we can
expand the denominator of the integrand in Eq.\ (\ref{dispersion}) as a power series in
$v k / \omega$.  Integrating by parts and solving the resulting
equation for $\omega^2$ gives
\be
\omega^2 = k^2 \left[ - \beta^2 + 3 \sigma^2 +
  O\left(\frac{\sigma^4}{\beta^2}\right) \right] \ ,
\label{omegaans}
\ee
where $\sigma^2 = \int dv v^2 F(v)$.
This equation has a solution with positive imaginary part,
$\omega_i>0$, consistent
with the assumption used in its derivation.  Therefore
the instability is generic, present for
any velocity distribution, as long as $\sigma \ll \beta$.
Equation (\ref{omegaans}) also suggests that the instability will be
removed when $\sigma$ gets large, as argued in Ref.\ \cite{Afshordi:2005ym},
since then the positive second term
in the square brackets will overcome the negative first term.
However, Eq.\ (\ref{omegaans}) is only valid in the regime $\sigma \ll
\beta$, and so to investigate stabilization we must use an alternative
method of computation.

Returning to the general dispersion relation (\ref{dispersion}), we obtain from
its real and imaginary parts the equations
\be
1 - k \beta^2 \int_{-\infty}^\infty dv \, \frac{ (\omega_r - k v )
  F'(v)}{(\omega_r - k v)^2 + \omega_i^2} =0 \ ,
\label{realpart}
\ee
and
\be
\int_{-\infty}^\infty dv \, \frac{
  F'(v)}{(\omega_r - k v)^2 + \omega_i^2} =0 \ .
\label{imagpart}
\ee
We now specialize to the Maxwellian velocity distribution
\be
F(v) = \frac{1}{\sqrt{2 \pi} \sigma} e^{ - \frac{ v^2}{2 \sigma^2}} \ .
\ee
For this case Eq.\ (\ref{imagpart}) has a unique solution for
$\omega_r$, namely $\omega_r =0$.  This can be seen by splitting the
integral into $v>0$ and $v<0$ contributions and substituting $v \to
-v$ in the $v<0$ term.  This yields
\be
0 =
\omega_r \int_0^\infty dv
\frac{ v^2 e^{- \frac{v^2}{2 \sigma^2}}}
{ \left[ (\omega_r - k v)^2 +
      \omega_i^2 \right] \left[ (\omega_r + k v)^2 +
      \omega_i^2 \right]} \ ,
\ee
and the result follows since the integrand is everywhere positive.
Equation (\ref{realpart}) now simplifies to
\be
\frac{\sigma^2}{\beta^2} = \int_{-\infty}^\infty d{\bar v}
\left(
\frac{
{\bar v}^2
}
{
{\bar v}^2 + \kappa^2
}
\right)
\, \frac{1}{\sqrt{2\pi}} e^{- \frac{{\bar v}^2}{2}} \ ,
\label{finale}
\ee
where we have defined ${\bar v} = v / \sigma$ and $\kappa = \omega_i /
(k \sigma)$.

Equation (\ref{finale})
determines $\kappa$ and thence $\omega_i$ as a function of $\sigma/\beta$.  When
$\kappa$ is large, expanding the factor in round brackets in the
integrand as a power series in ${\bar v}/\kappa$
gives the result (\ref{omegaans}) above.  As $\kappa$ decreases,
$\sigma/\beta$ increases, until as $\kappa \to 0$, $\sigma/\beta$
approaches the limiting value $\sigma/\beta =1$.  It can be seen that
there are no solutions to Eq.\ (\ref{finale}) with $\sigma/\beta>1$,
since the right hand side is a monotonic function of $\kappa^2$.
Therefore, whenever\footnote{Note that the critical value of velocity
  dispersion $\sigma=\beta$ is a factor of $\sqrt{3}$ larger than indicated by the
  approximate formula (\ref{omegaans}) which was used in Ref.\ \cite{Afshordi:2005ym}.}
\be
\sigma > \beta \ ,
\ee
there are no unstable modes with $\omega_i > 0$.  In other words, the
instability has been removed by the damping process associated with
the finite dispersion (Landau damping, also called free streaming).
The perturbation to the distribution function
is proportional to the integrand in Eq.\ (\ref{finale}).  Therefore
the nature of the damping is that interaction with the growing
mode moves some particles from velocities $v \sim \beta$ to larger
velocities, removing energy from the mode.

\section{Effective Newton's constant}
\label{sec:Geff}

In this appendix we derive the formula (\ref{Gformula}) for the effective Newton's
constant for the theory (\ref{action0}).

Since Newton's constant $G$ is dimensionful, we need to define the
system of units used in measurements of $G$, as our final result will
depend on the system of units chosen.  Here we are not concerned with
changes of units in the usual sense where the ratios between the old
and new standards of mass, length and time are constants, independent
of space and time.  Rather, we are concerned with changes in the
operational procedures for how units are defined, for which the ratios
between the old and new standards can vary with space and time, as
discussed by Dicke \cite{Dicke}.  Changes of units of this type
encompass conformal transformations of the metric, but can be more
general.  The necessity of specifying a system
of units before discussing the space-time variation of dimensionful
constants of nature is discussed in detail in Duff \cite{Duff}.

We choose to use systems of units which
are determined completely by physics in the
visible sector (which we label by ${\rm j}= b$ for baryonic), and are not
determined by
gravitational physics or by physics of the dark matter sector.
For example, this is true of SI units.  Alternatively one could take
one's standards of mass, length and time to be the mass, size, and
light travel time across a Hydrogen atom.  Any system of units of this
type will give rise to the formula (\ref{Gformula}) below for $G$.

We note that alternative choices are in principle possible.  For
example, one could have a small fiducial black hole, and one could transport
it adiabatically around the Universe to act as a standard.  Then one
could define the standard of mass to be the mass of the black hole,
while defining the standards of length and time in terms of Hydrogen
atoms as above.  In this system of units $G$ would vary in space and
time, but in a manner different to that described by the formula
(\ref{Gformula}) below, and Planck's constant $\hbar$ would also vary in
space and time.  Another possibility would be to define the units of
mass in terms of the mass of the dark matter particle; this would also
yield a different formula for $G$ whenever the dark matter coupling
function $\alpha_c(\phi)$ differs from that of the baryons
$\alpha_b(\phi)$.

We next discuss the quantities on which the formula for $G$ depends.  First,
for the theory (\ref{action0}) with several different matter sectors,
Newton's constant becomes a matrix whose i,j element governs the
strength of the gravitational interaction between sector ${\rm i}$ and
sector j.  Second, $G$ will depend on the scalar field
$\phi$, and through this dependence become a function of space and
time.  Here $\phi$ should be thought of as a background value of the
scalar field; gravitational interactions that act as perturbations
to this background, for which the perturbation to the
scalar field can be treated linearly, have a strength described by the
formula (\ref{Gformula}).  Finally, since the scalar contribution to
the gravitational force will in general have a Yukawa profile, $G$
will also depend on a spatial wave vector ${\bf k}$, which is measured
in the units discussed above.

Our formula for Newton's constant is
\be
G_{{\rm i}{\rm j}}(\phi,{\bf k}) = G e^{2 \alpha_b(\phi)} \left[ 1
  + \frac{2 \mpl^2 \alpha_{\rm i}^\prime(\phi) \alpha_{\rm j}^\prime(\phi)}{1 +
\frac{m_{\rm eff}^2(\phi,\rho)}{e^{2 \alpha_b(\phi)} {\bf k}^2 }
} \right] \ .
\label{Gformula}
\ee
Here on the right hand side, $G$ is a constant, related to $\mpl^2$ by
$G = 1 / (8 \pi \mpl^2)$, and the effective mass $m_{\rm eff}(\phi,\rho)$
is defined by
\be
m_{\rm eff}^2(\phi,\rho) = \frac{ \partial^2 V_{\rm
    eff}(\phi,\rho)}{\partial^2 \phi} \ ,
\label{meffgen}
\ee
where the effective potential is given by Eq.\ (\ref{phieqn}).  In the
adiabatic regime this effective mass reduces to the mass (\ref{meffdef}).
Several different aspects of this formula have appeared before in the
literature.  The overall prefactor of $e^{2\alpha_b}$ is well known, and
the formula for the case of one matter sector has been previously derived in
the context of cosmological perturbation theory \cite{Amendola:2004a}.
The $1$ in the square brackets describes the exchange of a graviton,
and the second term in the square brackets describes the exchange of a
scalar quantum, which couples to particles in the sector i with an
amplitude proportional to $\alpha_{\rm i}'(\phi)$.
The scalar coupling term
vanishes in the long wavelength limit $k \to 0$ if the scalar
field has a finite effective mass $m_{\rm eff}$.

We now turn to the derivation of the formula (\ref{Gformula}).
For the derivation it will be convenient to use a different system of
units, which are defined as follows.
The standards of mass, length and time are defined to be the mass,
size and light travel time across a small fiducial black hole, which
is transported adiabatically around the Universe to act as a
reference.  Equivalent units\footnote{By equivalent units we mean
  units that differ only by multiplication by constants.} can be
defined by demanding that
the speed of light $c$ and Planck's constant $\hbar$ be unity, and
that lengths and times are measured using the Einstein frame metric
$g_{ab}$. At the end of the derivation we will transform back to the
visible-sector-based units.

Starting with the action (\ref{action0}), we perform the following
steps.  We specialize the matter action in the jth sector to be that
of a collection of point particles with masses $m^{({\rm
    j})}_{\alpha_{\rm j}}$
and worldlines $x^{({\rm j})a}_{\alpha_{\rm j}}(\lambda)$:
\be
S_{\rm j}[e^{2 \alpha_{\rm j}(\phi)} g_{ab}, \Psi_{\rm j}] =
\sum_{\alpha_{\rm j}}
m^{({\rm j})}_{\alpha_{\rm j}} \int d\lambda \sqrt{ - e^{2 \alpha_{\rm
      j}} g_{ab}
  \frac{d x^{({\rm j})a}_{\alpha_{\rm j}}}{d\lambda}   \frac{d
    x^{({\rm j})b}_{\alpha_{\rm j}}}{d\lambda}} \ .
\ee
We specialize to the Newtonian limit, so that the Einstein-frame
metric $g_{ab}$ is of the form $- (1 + 2 \Phi) dt^2 + (1 - 2 \Phi)
d{\bf x}^2$.  We write the scalar field as $\phi + \delta \phi$, and
expand to quadratic order in $\delta \phi$ and $\Phi$.  We assume that there is a
background mass density $\rho$, so that the effective mass of the
scalar field is given by Eq.\ (\ref{meffgen}).
Finally, we compute from
the action the energy $E$ for a static configuration, up to an
additive constant.  This yields
\bea
E &=& \int d^3 x \left\{ \mpl^2 ({\bf \nabla} \Phi)^2 + \frac{1}{2} (
  {\bf \nabla} \delta \phi)^2 + \frac{1}{2} m_{\rm eff}^2 \delta
\phi^2 \right. \nn \\
&& + \sum_{\rm j} e^{\alpha_{\rm j}(\phi)} \left[ 1 + \Phi +
      \alpha_{\rm j}^\prime(\phi) \delta \phi \right] \rho_{\rm j}  \bigg\} \ ,
\eea
where
\be
\rho_{\rm j} = \sum_{\alpha_{\rm j}} m^{({\rm j})}_{\alpha_{\rm j}} \delta^3[ {\bf x} - {\bf
  x}^{({\rm j})}_{\alpha_{\rm j}} ] \ .
\ee
Now integrating out $\Phi$ and $\delta \phi$ yields
\bea
E &=& - \frac{1}{4 \mpl^2} \sum_{{\rm i},{\rm j}} \sum_{\alpha_1, \alpha_2, \ldots}
\sum_{\beta_1, \beta_2, \ldots} e^{\alpha_{\rm i}(\phi) + \alpha_{\rm j}(\phi)}
\frac{ m^{({\rm i})}_{\alpha_{\rm i}} m^{({\rm j})}_{\beta_{\rm
      j}}}{r^{{\rm i}{\rm j}}_{\alpha\beta}}
\nn \\
&& \times \left[ 1 + 2 \mpl^2
      \alpha_{\rm i}'(\phi) \alpha_{\rm j}'(\phi) e^{-
      m_{\rm eff} r^{{\rm i}{\rm j}}_{\alpha\beta} }\right] \ ,
\label{energy2}
\eea
where $r^{{\rm i}{\rm j}}_{\alpha\beta} = | {\bf x}^{({\rm
    i})}_{\alpha_{\rm i}} - {\bf x}^{({\rm j})}_{\beta_{\rm j}}|$.

Now we transform back to the visible sector or Jordan frame units.
The energy ${\tilde E}$, distances ${\tilde r}^{{\rm i}{\rm
    j}}_{\alpha\beta}$ and masses ${\tilde m}^{({\rm i})}_{\alpha_{\rm
    i}}$ in those units are given by
\bes
\bea
{\tilde E} &=& e^{-\alpha_b(\phi)} E,  \\
{\tilde r}^{{\rm i}{\rm j}}_{\alpha\beta} &=& e^{\alpha_b(\phi)}
r^{{\rm i}{\rm j}}_{\alpha\beta}, \\
{\tilde m}^{({\rm i})}_{\alpha_{\rm i}} &=& e^{-\alpha_b(\phi)} \left[
  e^{\alpha_{\rm i}(\phi)} m^{({\rm i})}_{\alpha_{\rm i}} \right] \ ,
\label{jmass}
\eea
\ees
where the factor in square brackets on the right hand side of Eq.\
(\ref{jmass}) is the mass as measured in the Einstein-frame units.
Substituting these unit transformations into the energy expression
(\ref{energy2}) and transforming from position space to momentum space
yields an expression for the energy from which we can
finally read off the effective Newton's constant (\ref{Gformula}).

\end{document}